\documentclass[12pt]{amsart}
\usepackage[T1]{fontenc}
\usepackage[utf8]{inputenc}
\usepackage[final]{microtype}
\usepackage{lmodern}

\usepackage[margin=1in]{geometry}

\usepackage{amssymb,mathrsfs}
\usepackage{dsfont}
\usepackage{esint}
\usepackage{slashed}
\usepackage{faktor}
\usepackage{todonotes}

\usepackage{chngcntr}
\counterwithin{equation}{section}

\usepackage{amsthm}
\usepackage{thmtools, thm-restate}
\declaretheorem[name=Theorem]{theorem}
\counterwithin{theorem}{section}
\declaretheorem[name=Proposition, numberlike=theorem]{prop}

\declaretheorem[name=Definition, numberlike=theorem]{defn}
\declaretheorem[name=Lemma, numberlike=theorem]{lemma}
\declaretheorem[name=Corollary, numberlike=theorem]{cor}

\declaretheorem[name=Remark, numberlike=theorem, style=remark]{rmk}

\usepackage{mathtools}
\mathtoolsset{showonlyrefs}

\usepackage{csquotes}

\usepackage[final]{hyperref}

\usepackage{tikz-cd}

\tikzset{
	pf/.style={commutative diagrams/.cd, every arrow, every label},
	surj/.style=commutative diagrams/two heads,
	inj/.style=commutative diagrams/hook,
	gl/.style=commutative diagrams/equal,
	mat/.style={matrix of math nodes, commutative diagrams/.cd, every cell},
	dr/.style={matrix of math nodes, commutative diagrams/.cd, every cell, column sep=small},
	seq/.style={matrix of math nodes, commutative diagrams/.cd, every cell,column sep=small}
}

\newenvironment{diag*}
	{\[\begin{tikzpicture}[commutative diagrams/.cd, every diagram, baseline=(current bounding box.center)]}
	{\end{tikzpicture}\]\ignorespacesafterend}

\newenvironment{diag}
	{\begin{equation}\begin{tikzpicture}[commutative diagrams/.cd, every diagram, baseline=(current bounding box.center)]}
	{\end{tikzpicture}\end{equation}\ignorespacesafterend}

\newcommand{\Action}{\YMHD}
\newcommand{\Conn}{\mathfrak{A}} 
\newcommand{\dd}{\mathop{}\!\mathrm{d}}
\newcommand{\D}{\slashed{D}}
\newcommand{\dt}[1]{\frac{\dd}{\dd{t}}\Big|_{_{t=#1}}}
\newcommand{\DA}{A_D}
\newcommand{\fiber}{N} 
\newcommand{\FB}{\mathcal{N}} 
\newcommand{\fbm}{\mathscr{G}} 
\newcommand{\fm}{h} 
\newcommand{\g}{\mathfrak{g}}
\newcommand{\Gaug}{\mathscr{D}} 
\newcommand{\HB}{\mathcal{H}} 

\newcommand{\LC}{\overline{\nabla}} 
\newcommand{\p}{\partial}
\newcommand{\pd}{\slashed{\partial}}
\newcommand{\Qpsi}{\mathcal{Q}} 
\newcommand{\R}{\mathbb{R}}
\newcommand{\VB}{\mathcal{V}} 
\newcommand{\vbm}{\bar{h}} 
\newcommand{\VC}{\mathcal{R}^\VB} 
\newcommand{\potential}{W}
\newcommand{\YM}{A_{YM}}
\newcommand{\YMHD}{\mathcal{A}}
\newcommand{\II}{\mathrm{I\hspace*{-0.4ex}I}} 

\DeclareMathOperator{\Ad}{Ad}
\DeclareMathOperator{\ad}{ad}

\DeclareMathOperator{\diverg}{div}

\DeclareMathOperator{\dom}{Dom}
\DeclareMathOperator{\dv}{\dd{vol}}
\DeclareMathOperator{\dx}{\dd{x}}
\DeclareMathOperator{\End}{End}

\DeclareMathOperator{\Hom}{Hom}
\DeclareMathOperator{\hor}{hor}
\DeclareMathOperator{\id}{Id}
\DeclareMathOperator{\Isom}{Isom}
\DeclareMathOperator{\Ker}{Ker}
\DeclareMathOperator{\Osc}{Osc}
\DeclareMathOperator{\pr}{pr}
\DeclareMathOperator{\Spin}{Spin}

\DeclareMathOperator{\supp}{supp}

\DeclareMathOperator{\tr}{Tr}
\DeclareMathOperator{\ver}{ver}
\DeclareMathOperator{\Vol}{Vol}

\begin{document}

\title{Geometric analysis of the Yang--Mills--Higgs--Dirac~model}

\date{\today}

\author{Jürgen Jost}
\address{
	Max Planck Institute for Mathematics in the Sciences\\
	Inselstr. 22--26\\
	04103 Leipzig\\
	Germany\\
\newline
Santa Fe Institute for the Sciences of Complexity\\
Santa Fe, NM 87501\\
USA
}
\email{jjost@mis.mpg.de}

\author{Enno Keßler}
\address{%
	Harvard University,
	Center of Mathematical Sciences and Applications,
	20 Garden Street,
	Cambridge, MA 02138,
	USA
}
\curraddr{%
	Max-Planck-Institut für Mathematik\\
	Vivatsgasse 7\\
	53111 Bonn\\
	Germany
}
\email{kessler@mpim-bonn.mpg.de}
\thanks{Enno Keßler was supported by a Deutsche Forschungsgemeinschaft Research Fellowship, KE2324/1--1.}

\author{Ruijun Wu}
\address{
	SISSA\\
	Via Bonomea 265\\
	34136 Trieste\\
	Italy
}
\curraddr{%
	School of Mathematics and Statistics\\
	Beijing Institute of Technology\\
	Zhongguancun South Street No. 5\\
	100081 Beijing\\
	P.R.China
}
\email{ruijun.wu@bit.edu.cn}
\thanks{Ruijun Wu was supported by the Centro di Ricerca Matematica ‘Ennio de Giorgi’ and by SISSA}

\author{Miaomiao Zhu}
\address{
	School of Mathematical Sciences\\
	Shanghai Jiao Tong University\\
	Dongchuan Road 800\\
	200240 Shanghai\\
	P.R.China
}
\email{mizhu@sjtu.edu.cn}
\thanks{Miaomiao Zhu was partially supported by Innovation Program of Shanghai Municipal Education Commission (No. 2021-01-07-00-02-E00087), National Natural Science Foundation of China (No. 12171314) and Shanghai Frontier Science Center of Modern Analysis.}

\begin{abstract}
	The harmonic sections of the Kaluza--Klein model can be seen as a variant of harmonic maps with additional gauge symmetry.
	Geometrically, they are realized as sections of a fiber bundle associated to a principal bundle with a connection.
	In this paper, we investigate geometric and analytic aspects of a model that combines the Kaluza--Klein model with the Yang--Mills action and a Dirac action for twisted spinors.
	In dimension two we show that weak solutions of the Euler--Lagrange system are smooth.
	For a sequence of approximate solutions on surfaces with uniformly bounded energies we obtain compactness modulo bubbles, namely, energy identities and the no-neck property hold.
\end{abstract}

\maketitle

{\footnotesize

\medskip

}


\section{Introduction}
In this article we study geometric and analytic properties of a gauged non-linear sigma model that combines the theory of harmonic sections with the Yang--Mills action and the Dirac action for vector-valued spinors.
Mathematically this yields an equivariant extension of the theory of the theory of Dirac-harmonic maps and is also motivated as a simplification of supersymmetric Yang--Mills theory from physics, as well as a twisted fiber bundle version of (anti-)self-dual equations on Riemann surfaces.

Investigating the rich and subtle mathematical structure of quantum field theory (QFT) is important for physics and mathematics alike.
In fact, models from QFT have lead to a host of powerful geometric invariants.
In particular, Donaldson could construct powerful invariants for differentiable 4-manifolds from solution spaces of anti-self-dual Yang--Mills connections, and later, Seiberg and Witten derived simpler invariants also from the Yang--Mills functional.
The Gromov--Witten invariants are fundamental in symplectic geometry, to name just the most famous and powerful such invariants.

The Yang--Mills functional evaluates the $L^2$-norm of the curvature of a connection on a principal bundle.
Such a connection arises as a gauge field in QFT.\@
The first gauge theory was proposed by Hermann Weyl, in order to unify electromagnetism with gravity.
The gauge group was the abelian group $U(1)$.
While this was not successful as a physical theory, it inspired Yang and Mills to develop gauge theories with non-abelian gauge groups.
Yang--Mills--Higgs theory couples the connection from Yang--Mills theory with a section of an associated bundle of the principal bundle, the Higgs field.
These theories constitute the basis of the Standard Model of elementary particle physics that unifies the electromagnetic, weak and strong forces.
The gauge group here is $SU(3)\times SU(2)\times U(1)$, but mathematically, one can work with any compact linear group.
Thus, also grand unified theories with gauge groups like $SU(5)$ have been proposed.
The gauge fields, however, constitute only half of the fields of QFT, the bosonic ones.
The other fields are the fermionic matter fields.
They are mathematically represented by spinors, and the action is of Dirac type.
These two types of fields are combined in supersymmetric Yang--Mills theory.
The action functional includes commuting gauge fields and anti-commuting matter fields, and supersymmetry converts one type of field into the other, while leaving the action invariant.
The supersymmetric Yang--Mills action is mathematically very rich.
In order to develop tools for its mathematical analysis and to explore its geometric consequences, it has been found expedient to work with simplified versions.
For instance, the Seiberg-Witten invariants arise from a reduced version of super Yang--Mills.
Perhaps the simplest action functional that still captures the essential mathematical aspects behind super Yang--Mills is the nonlinear supersymmetric sigma model, see for instance~\cite[Chapter 6]{deligne1999supersolutions}.
Here, the gauge connection is replaced by a map into some Riemannian manifold (a sphere in the original model, but mathematically, one can take any Riemannian manifold).
The action functional for that map is the Dirichlet action%
\footnote{In the mathematical literature, this is usually called an energy instead of an action; in fact, we shall use some energies below for auxiliary purposes in our analysis.}%
, and its critical points are known as harmonic maps in the mathematical literature.
The matter field becomes a spinor field along the map, and the critical points solve a nonlinear Dirac equation.
For the details of the algebraic and geometric structure of this action functional, we refer to the systematic investigation~\cite{kessler2019super}.

From a semiclassical perspective, one would like to study the critical points of the action functional.
They are solutions of certain partial differential equations (PDEs), the Euler--Lagrange equations for the functional.
Here, a new mathematical difficulty arises.
The fermionic fields are anti-commuting, and therefore, they are not amenable to regularity theory for solutions of partial differential equations, because that theory works with analytical inequalities, and these are meaningful only for commuting (real-valued) fields.
Therefore, in~\cite{chen2006dirac}, a variant of the functional has been constructed that works with commuting fields only.
That is, the spinor fields also become commuting fields.
This is achieved by changing the Clifford algebra for the representation of the spin group.
By that construction, supersymmetry between the fields is lost, but all other symmetries, in particular conformal symmetry, are preserved, and the analytical power of PDE regularity theory is gained.

In this paper we generalize those semiclassical methods to include more fields of the standard model.
A similar model was also proposed by Witten~\cite{witten1988topological}, where he briefly indicated the possibility of a generalization of sigma models in this direction.
Here we develop an even more general geometric setup that couples Yang--Mills theory with the Higgs-field and a Dirac action for commuting twisted spinors.
The Higgs field whose physical role consists in assigning masses to other fields is of particular geometric significance and requires to consider harmonic sections in the Kaluza--Klein geometry.
Coupling the commuting spinors to the Kaluza--Klein geometry requires taking tensor products with the vertical part of the fiber bundle.
The resulting Euler--Lagrange equations which we derive in detail couple the Yang--Mills--Higgs equations with a Dirac equation for the commuting spinors.
Furthermore, we start to explore analytic properties of the Euler--Lagrange equations:
if the dimension of the domain is two, we show regularity of its solutions up to gauge transformations and study the bubbling behavior.
This is more or less standard in analysis, so we only briefly sketch the results.
We develop a general abstract model, which includes as examples those known in the standard product case where the fibers are Kähler manifolds or Calabi--Yau manifolds.
It remains a task for the future to explore new examples.
In any case, this fiber bundle construction would potentially lead to a different perspective for the equivariant version of supersymmetric Yang--Mills equation, or gauged Dirac-harmonic maps, since the equations take care of the global nontriviality of the fiber bundles.

Let us now describe the geometric structure in more detail.
Given a \(G\)-principal fiber bundle~\(P\) over the manifold~\(M\) and a left~$G$-manifold~$(\fiber, \fm)$, one can construct the associated fiber bundle~$\FB=P\times_G\fiber$ over \(M\).
A principal connection~$\omega$ on~$P$ induces an associated connection on~$\FB$, in particular a splitting \(T\FB=\mathcal{H}\oplus\mathcal{V}\) of the tangent bundle in a horizontal and a vertical part.
Kaluza--Klein theory constructs a bundle metric~$\fbm$, turning~$(\FB,\fbm)$ into a Riemannian manifold.
While the action on the connection is given by the Yang--Mills functional, the Higgs energy of sections \(\phi\colon M\to\FB\) is not the full Dirichlet energy, since the latter is not compatible with the variation in the space of sections of \(\FB\).
Rather, we should restrict it to the vertical part \(\dd^\VB\phi\) of its differential.
Similarly, we define a vertical Dirac-operator~\(\D\) acting on the spinors~\(\psi\in\Gamma(S\otimes \phi^*\VB)\).
Putting the pieces together, the action of the Yang--Mills--Higgs--Dirac functional is then given by
\begin{equation}
	\Action(\omega,\phi,\psi)
	=\int_M |F(\omega)|^2 + |\dd^\VB\phi|^2+ \left<\psi,\D\psi\right> \dv_g,
\end{equation}
where~$F(\omega)$ is the curvature of the principal connection~$\omega$.

\begin{restatable*}{theorem}{ELEquations}%
\label{thm:ELEquations}
	The critical points of the coupled action functional \(\Action(\omega, \phi, \psi)\) are the solution of the following Euler--Lagrange equations
	\begin{equation}\label{eq:EL-global}
		\begin{aligned}
			&D_\omega^* F+ \dd\bar{\mu}^*_\phi(\dd^\VB\phi)+ \Qpsi(\phi,\psi)=0,\\
			&\tau^\VB(\phi)-\frac{1}{2}\VC(\phi,\psi)=0, \\
			&\D\psi=0.
		\end{aligned}
	\end{equation}
\end{restatable*}
The terms \(\dd\bar{\mu}^*_\phi(\dd^\VB\phi)\) and \(\Qpsi(\phi, \psi)\) describe the infinitesimal dependence of \(|\dd^\VB\phi|^2\) and \(\left<\psi, \D\psi\right>\) on \(\omega\), respectively.
The vertical tension field \(\tau^\VB(\phi)\) is a differential operator of order two and \(\VC(\phi, \psi)\) is a contraction of the Riemannian curvature of \(\fbm\).
Up to the choice of a gauge, for instance the Coulomb gauge,~\eqref{eq:EL-global} is locally an elliptic system.

For the geometric constructions, there exists some prior work on which we build.
The Yang--Mills--Higgs theory has been analyzed from a mathematical perspective, viewing the Higgs field as a natural generalization of harmonic maps to fiber bundles as C.\ M.\ Wood noticed in his work~\cite{wood1987harmonic, wood1990existence} on harmonic sections.
David Betounes has clarified that the right geometric setup for Yang--Mills--Higgs theory is given by a Riemannian variant of Kaluza--Klein geometry, see~\cite{betounes1989geometry, betounes1991kaluza, betounes2004mathematical}.
The Yang--Mills--Higgs functional is also investigated under the name of gauged harmonic maps by Lin--Yang~\cite{LinYang2003Gauged}.
Thomas H. Parker~\cite{parker1982gauge} has worked on the unification of Yang--Mills theory with spinors in the case of linear fiber bundles.
We point out that the geometric setup developed here works, more generally, for non-linear fibers and integrates twisted spinors in the Kaluza--Klein geometry.

Analytical properties of the system~\eqref{eq:EL-global} depend heavily on the conformal symmetry of the system, hence on the dimension of the domain.
While Yang--Mills theory is richest in dimension four, the theory of harmonic maps, that is, the Higgs-field, encounters non-resolvable singularities already in dimension three.
Consequently, sigma models are usually studied in dimension one or two, and we will also restrict our attention here to the case of a two-dimensional domain.
Atiyah--Bott~\cite{atiyah1983}, and N. Hitchin~\cite{hitchin1987selfdual},
have demonstrated, Yang--Mills theory also leads to geometric and topological insight in dimension two.
Based on Karen Uhlenbeck's Coulomb gauge theorem and Tristan Rivière's regularity theory, we obtain the full regularity of weak solutions on surfaces.
\begin{restatable*}{theorem}{RegularityOfWeakSolution}%
\label{thm:regularity of weak solution}
	Let~$(M,g)$ be a closed Riemann surface.
	Let~$(\omega,\phi,\psi)$ be a weak solution of~\eqref{eq:EL-global}.
	Then there is a gauge transformation~$\varphi\in \Gaug^{2,2}$ such that~$(\varphi^*\omega, \varphi^*(\phi),\varphi^*(\psi))$ is smooth.
\end{restatable*}

The bubbling phenomenon of harmonic maps is also special to the case of two dimensions because of the conformal symmetry of the Dirichlet energy part.
However, the coupled Yang--Mills--Higgs--Dirac action is not conformally invariant because the Yang--Mills term is not conformally invariant in dimension two.
Hence, for sequences of approximating solutions and a suitable notion of energy, the energy only concentrates at the blow-up points of the sections and the connection does not contribute to energy concentration.
After establishing the necessary small energy regularity, we show that the limit objects are bubble trees of Dirac-harmonic maps with trivial principal bundle and connection.
\begin{restatable*}{theorem}{EnergyIdentitiesNoNeck}%
\label{thm:energy identities and no-neck}
	Let~$(\omega_k,\phi_k,\psi_k)$ be a sequence of approximating solutions to the Euler--Lagrange system~\eqref{eq:EL-global} with uniformly bounded energies.
	Then up to extraction of a subsequence the sequence of approximating solutions converges weakly to a smooth solution~$(\omega_\infty, \phi_\infty,\psi_\infty)$ of~\eqref{eq:EL-global}.

	Furthermore, there is a finite set~$\mathcal{S}_1=\{ x_1, \dotsc, x_I\}\subset M$ such that the convergence is strong on any compact subset of~$M\setminus \mathcal{S}_1$.
	For each~$x_i\in\mathcal{S}_1$ there exists a finite collection of Dirac-harmonic spheres~$(\sigma^l_i, \xi^l_i)$ from~$\mathbb{S}^2$ into~$\fiber$ for~$1\le l\le L_i <\infty$, such that the energy identities
	\begin{align}
		\lim_{k\to\infty} \YM(\omega_k)&=\YM(\omega_\infty), \\
		\lim_{k\to\infty} E(\phi_k) &=E(\phi_\infty)+\sum_{i=1}^I \sum_{l=1}^{L_i} E(\sigma^l_i), \\
		\lim_{k\to\infty} E(\psi_k) &=E(\psi_\infty)+\sum_{i=1}^I \sum_{l=1}^{L_i} E(\xi^l_i),
	\end{align}
	and the no-neck property hold, that is, \(\phi_\infty(M)\cup\left(\cup_{i,l}\sigma^l_i(\mathbb{S}^2)\right)\) is connected.
	The principal bundle and connection on the bubbles \((\sigma^l_i, \xi^l_i)\) are trivial.
\end{restatable*}
It remains to be seen if and how the resulting bubble trees can be seen as compactification points of a set of first order equations.
For the treatment of twisted holomorphic maps, a set of first order equations that minimize a Yang--Mills--Higgs without spinors, we refer to~\cite{riera2009compactification}.
When the complex structure of the domain varies and degenerates to some surface with nodes, then the connection part plays a special role in the analysis of the degenerating region and a new phenomenon occurs, see~\cite{riera2009compactification, song2016convergence}.
We expect that a similar phenomenon can be explored in the case of  Yang--Mills--Higgs--Dirac model.

The article is organized as follows.
In Section~\ref{sec:GeometryOfTheModel} we give a detailed geometric setup of the Yang--Mills--Higgs--Dirac model.
For the convenience of the readers and in order to fix the notation, we recall the Kaluza--Klein geometry of general fiber bundles over Riemannian manifolds and the theory of harmonic sections.
Afterwards we formulate the coupled model and derive its equations of motions for all fields and obtain local expressions.
In Section~\ref{Sec:Regularity} we derive the regularity of weak solutions of the equations of motion up to a gauge transformations for the case where the dimension is two.
In Section~\ref{sec:SmallEnergyRegularity} we obtain the small energy regularity as a first step towards understanding the limiting behavior of critical points.
Section~\ref{sec:BlowUpAnalysis} treats the bubbling of the Yang--Mills--Higgs--Dirac model by proving that the energy of the connection does not concentrate and subsequent reduction to the case of Dirac-harmonic maps.
In the second part we focus on the case where the domain is a closed Riemann surface.
With the help of the tools from Yang--Mills and harmonic map theory, we obtain the regularity of weak solutions and the energy identities and no-neck properties for the approximate sequences.

\section{Geometry of the Model}%
\label{sec:GeometryOfTheModel}
In this section we describe the geometric setup of the Yang--Mills--Higgs--Dirac model:
We will give an explicit description of the construction of the Kaluza--Klein metric on the associated fiber bundle \(\FB\) from a connection \(\omega\) on the principal bundle \(P\to M\) and the Riemannian metrics on \(M\) and the fiber \(\fiber\).
Sections \(\phi\) of \(\FB=P\times_G \fiber\) are called harmonic sections if they are critical points of a suitable generalization of the Dirichlet action: the Higgs action.
Afterwards we couple the Higgs action with the Yang--Mills action and investigate the dependence of the Higgs action on the connection.
In the last step we build the Yang--Mills--Higgs--Dirac action by adding the Dirac action for vertical twisted spinors.

Several aspects and special cases of this coupled action have been studied before:
The Higgs action has been introduced in~\cite{wood1987harmonic, wood1990existence} as an equivariant generalization of harmonic maps and has been studied more recently in~\cite{LE} under the aspect of harmonicity of geometric structures.
The Kaluza--Klein metric and its geometry has been investigated in~\cite{betounes1991kaluza, betounes2004mathematical}.
In~\cite{parker1982gauge} the coupling of the Yang--Mills equation with Laplace equations and Dirac equations was analyzed on associated vector bundles over four-dimensional Riemannian manifolds.
Our work combines the above approaches to a geometric setup for the coupled Yang--Mills--Higgs--Dirac action in arbitrary dimension of the base manifold \(M\) and for general non-linear fiber \(N\).
The main technical achievement of this section is to obtain the precise dependence of Kaluza--Klein geometry and, subsequently, of the coupled action on the connection.

\subsection{Kaluza--Klein metric on associated fiber bundles}
When the structure group of a principal fiber bundle acts on a vector space one obtains an associated vector bundle.
Here we describe the necessary generalization to the non-linear case where the fiber is a general Riemannian manifold.
We will see that a connection on the principal bundle allows us to combine the metric on the base manifold and the fiber manifold to a Riemannian metric on the associated fiber bundle, called Kaluza--Klein metric in~\cite{betounes2004mathematical}.

Let~$M$ be an $m$-dimensional oriented closed manifold with a Riemannian metric~$g$, and~$G$ a compact Lie group with Lie algebra~$\g$.
In particular, being compact and hence finite-dimensional,~$G$ is isomorphic to a matrix group.
Suppose, $P=P(M,G,\pi,\Psi)$ is a principal $G$-bundle over~$M$, where~$\pi\colon P\to M$ denotes the projection and
\begin{align}
	\Psi\colon P\times G&\to P, &
	\Psi(p,a)&\equiv\Psi_a(p)\equiv \Psi_p(a),
\end{align}
denotes a free right~$G$-action.
Further assume that~$(\fiber,\fm)$ is a left~$G$-manifold; that is,
\begin{align}
	\mu\colon G\times \fiber&\to \fiber, &
	\mu(a,y)&\equiv \mu_a(y)\equiv \mu_y(a),
\end{align}
is a left action with~$\mu(G)\subset \Isom(\fiber,\fm)$.
Then~$G$ acts on the product~$P\times N$ from the right freely:
\begin{align}
	(\Psi\times \check{\mu})\colon (P\times \fiber)\times G&\to P\times\fiber \\
	((p,y),a)&\mapsto (\Psi_a(p),\mu_{a^{-1}}(y)).
\end{align}
For further reference we abbreviate \(\check{\mu}_a \coloneqq \mu_{a^{-1}}\).
The orbit space
\begin{equation}
	\FB\coloneqq \faktor{(P\times \fiber)}{G} \equiv P\times_G \fiber
\end{equation}
is a smooth manifold; denote the quotient map by~$\iota\colon P\times N\to \FB$, where~$\iota(p,y)=[p,y]$.
Note that this is a principal fiber bundle over~$\FB$ with fiber~$G$.
As \(G\) acts fiberwisely on \(P\), there is a unique map~$\rho\colon \FB\to M$ s.t.\ the following diagram commutes:
\begin{diag}
	\matrix[mat, column sep=small](m)
	{
		P\times \fiber & \faktor{\left(P\times \fiber\right)}{G}=P\times_G \fiber \\
		P       & M \\
	};
	\path[pf]
		(m-1-1) edge node{$\iota$} (m-1-2)
						edge node{$\pr_1$} (m-2-1)
		(m-2-1) edge node{$\pi$}  (m-2-2)
		(m-1-2) edge [dashed] node{$\exists ! \rho$} (m-2-2)
		;
\end{diag}
It is well-known that~$\rho\colon \FB\to M$ is a fiber bundle with fiber space~$\fiber$.
The embedding of the fiber is given by the insertion map:
For any~$x\in M$ and any~$p\in P$ with~$\pi(p)=x$, the insertion map
\begin{equation}
	\begin{split}
		\iota_p\colon \fiber &\to \FB_x=\rho^{-1}(x), \\
		y &\mapsto [p,y]
	\end{split}
\end{equation}
is a diffeomorphism.
A different choice of the point \(p\) yields an embedding differing by an automorphisms of~$\fiber$.

The differential \(\dd\rho\) of the projection \(\rho\colon \FB\to M\) yields a short exact sequence of vector bundles over the fiber bundle~$\FB$:
\begin{diag}\label{seq:SES over P}
	\matrix[mat, column sep=normal](m)
	{
		0 & \VB\equiv \Ker(\dd\rho) & T\FB  & \rho^*(TM) & 0\\
			&										& \FB  & &  \\
	};
	\path[pf]
		(m-1-1) edge (m-1-2)
		(m-1-2) edge node[swap]{\(\pi^\VB\)} (m-2-3)
						edge[commutative diagrams/hook] (m-1-3)
		(m-1-3) edge (m-2-3)
						edge node{$\dd\rho$} (m-1-4)
		(m-1-4) edge (m-2-3)
		(m-1-4) edge (m-1-5)
	;
\end{diag}
We call~$\pi^\VB\colon \VB\to \FB$ the \emph{vertical bundle} over~$\FB$, with fibers given by tangent spaces of~$N$.

Analogously, the vertical bundle \(VP\subset TP\) of a principal bundle \(P\) is defined by \(VP=\Ker{\dd\pi}\) and can be shown to be trivial: \(VP=P\times\g\).
A \emph{principal connection} is a \(G\)-invariant splitting \(TP=VP\oplus HP\), where the horizontal bundle \(HP\) is isomorphic to \(\pi^*TP\).
Such a principal connection can be given by a \(G\)-equivariant~$\g$-valued one form \(\omega\in\Omega^1(P, \g)\) such that, under the above trivialization, \(\omega((p,\mathfrak{a}))=\mathfrak{a}\in\g\) for all \((p,\mathfrak{a})\in VP=P\times\g\).
The kernel of~$\omega$ is the horizontal distribution~$HP$.
For more details on connections in principal bundles, see~\cite{rudolph2017differential}.

The principal connection~\(\omega\) induces an Ehresmann connection~$\sigma$ on the associated bundle~$\FB$ by specifying a horizontal distribution~$\HB$ complementary to~$\VB$ in~$T\FB$:
\begin{equation}
	\HB_{|_{[p,y]}}\coloneqq \dd\iota_y({(HP)}_p).
\end{equation}
Here, \(\iota_y\colon P\to\FB\) is the map that arises from \(\iota\colon P\times N \to\FB\) by restricting to \(y\in N\) and \(\dd\iota_y\) its tangent map.
The equivariance of~$\omega$ guarantees that distribution \(\HB\) is well-defined.
By construction we have $\HB_{|_{[p,y]}}\cong T_{x}M$ where~$x=\rho([p,y])=\pi(p)$, namely~\(\HB\cong \rho^*TM\).
More explicitly, if~$\tilde{X}$ is a lift of~$X\in\Gamma(TM)$ with horizontal part~$\hor\tilde{X}$, then the isomorphism
\begin{equation}
		\sigma\colon \rho^*TM\xrightarrow{\cong} \HB\subset T\FB
\end{equation}
is then given by
\begin{align}
	{\sigma(\rho^*X)}_{|_{[p,y]}}= \dd\iota_y(\hor\tilde{X}).
\end{align}
Since~\(\rho\circ\iota_y =\pi\) for all~\(y\in N\), the short exact sequence~\eqref{seq:SES over P} splits via~$\sigma$:~\(\dd\rho\circ\sigma = \id_{\rho^*TM}\).
Thus~\(T\FB=\VB\oplus\HB\) where the projectors on the horizontal and vertical bundles are given by
\begin{align}
	\hor&=\sigma\circ \dd\rho, &
	\ver&=(\mathds{1}-\sigma\circ\dd\rho).
\end{align}
In particular, the map \(\sigma\) defines a connection on \(\FB\).

The embeddings~$\iota_p\colon N\to \FB$ of fibers~\(p\in P\) induce a Riemannian metric~$\vbm$ on~$\pi^\VB$:
\begin{equation}
	\vbm_{|_{[p,y]}}\coloneqq {(\iota_p^{-1})}^* (\fm_{|_y}), \qquad
	\forall [p,y]\in\FB,
\end{equation}
which is well-defined since~$\mu(G)\subset\Isom(\fiber,\fm)$.
Together with the splitting \(T\FB=\VB\oplus\HB\) from the connection we can define the \emph{Kaluza--Klein metric}
\begin{equation}
	\fbm(X, Y)= \vbm\left(\ver X, \ver Y\right) + g_\rho(\hor X, \hor Y),
	\qquad
	\forall X,Y\in\Gamma(T\FB).
\end{equation}
Here, \(g_\rho\) is the pull-back metric on~\(\HB=\rho^*TM\) obtained from~\(g\) on~\(TM\) via~\(\rho\).
As a Riemannian manifold,~$(\FB,\fbm)$ admits a unique Levi-Civita connection~$\LC$.

With respect to the Kaluza--Klein metric~$\fbm$, the fibration~$\rho\colon \FB\to M$ has totally geodesic fibers.
As a consequence, for vector fields \(Y\) and \(Z\) on \(N\), the Levi-Civita covariant derivative of the local vertical vector fields~$\dd\iota_p(Y)$, $\dd\iota_p(Z)$ on~$\FB$ is given by
\begin{equation}
	\LC_{\dd\iota_p(Y)}\dd\iota_p(Z)=\dd\iota_p(\nabla^h_Y Z).
\end{equation}
In other words, for vertical vector fields~\(W\) and~\(V\) the covariant derivative~\(\LC_{W}V\) is again vertical.
It follows that also for horizontal vector fields~\(H\) the field~\(\LC_W H\) is horizontal and~\(\LC_W\hor = \LC_W\ver =0\).
Further properties of the Levi-Civita connection~\(\LC\) have been investigated in~\cite{betounes2004mathematical}.

\subsection{Harmonic sections}
Harmonic sections of the associated fiber bundle \(\FB\) are critical points of an action analogous to the Dirichlet action to be defined in~\eqref{eq:energy of sections} below.

Notice that, in general, there are topological obstructions to the existence of sections of fiber bundles, see~\cite[Section 29]{steenrod1951topology}.
For instance, the case of dimension two, which we are mainly interested in, is unobstructed if the second homotopy group of the fiber vanishes.
Sections which are at least differentiable once can then be turned into smooth sections by local approximation.

From now on, we assume the existence of a smooth section~\(\phi\in\Gamma(\FB)\).
The pull back of~\eqref{seq:SES over P} along \(\phi\) yields short exact sequence of vector bundles over~$M$:
\begin{equation}
	\begin{tikzcd}
		0\arrow{r} & \phi^*\VB\arrow{r}\arrow{dr} & \phi^*T\FB \arrow{r}{\dd\rho}\arrow{d} & TM \arrow{dl}{\pi_M}\arrow{r} & 0 \\
		&& M &&
	\end{tikzcd}
\end{equation}
The horizontal part of the differential~$\dd\phi\in \Gamma(T^*M \otimes \phi^*T\FB)$ is the identity \(\mathds{1}_{TM}\) and hence has constant length~\(\sqrt{m}\).
The vertical part~$\dd^\VB\phi\equiv \ver\dd\phi\in\Gamma(T^*M\otimes \phi^*\VB)$ encodes the essential geometric information contained in the gradient of the section.
Therefore we consider the effective Dirichlet energy of the section defined by
\begin{equation}\label{eq:energy of sections}
	E(\phi;\sigma)\coloneqq \int_M |\dd^\VB\phi|^2_{g^\vee\otimes \fbm} \dv_g
\end{equation}
where~$g^\vee$ denotes the dual metric on the cotangent bundle~$T^*M$.
As the decomposition~$T\FB=\HB\oplus \VB$ is orthogonal with respect to~$\fbm$ it holds
\begin{equation}
	\int_M |\dd\phi|^2_{g^\vee\otimes \fbm}\dv_g = E(\phi;\sigma)+\dim(M)\cdot \Vol(M).
\end{equation}

The Dirichlet energy functional~\eqref{eq:energy of sections} can be defined on the space of~$W^{1,2}$-sections.
Its critical points are known as \emph{harmonic sections}, see~\cite{wood1987harmonic, wood1990existence}.
The Euler--Lagrange equation of~\eqref{eq:energy of sections} is derived as follows.
Let~$\phi\in\Gamma(\FB)$ and take a variation~$(\phi_t)$ of~$\phi$ in the space of~$W^{1,2}$-sections.
Thus the variational field
\begin{equation}
	V=\dt{0}\phi_t
\end{equation}
is vertical.
Direct calculation shows that
\begin{equation}\label{eq:variation-vertical harmpnic maps}
	\begin{split}
		\frac{\dd}{\dd t} E(\phi_t)
		={}& \left.\frac{\dd}{\dd t}\right|_{t=o} \int_M \left<\dd^\VB\phi_t(e_\alpha), \dd^\VB\phi_t(e_\alpha)\right>\dv_g \\
		={}& -2\int_M \left< V, \LC_{e_\alpha}^{\phi^*T\FB}\dd^\VB\phi(e_\alpha) + \left(\diverg_g e_\alpha\right)\dd^\VB\phi(e_\alpha)\right>\dv_g
	\end{split}
\end{equation}
Thus the critical points of~\eqref{eq:energy of sections} satisfies the equation
\begin{equation}\label{eq:harmonic section equation-global}
	\tau^\VB(\phi)
	\coloneqq	\ver\LC_{e_\alpha}^{\phi^*T\FB}\dd^\VB\phi(e_\alpha) + \left(\diverg_g e_\alpha\right)\dd^\VB\phi(e_\alpha)
	=0,
\end{equation}
where~$\diverg_g(e_\alpha)\equiv\sum_{\beta} \left<\nabla_{e_\beta} e_\alpha, e_\beta\right>$.
This tensor~$\tau^\VB(\phi)$ is called the \emph{vertical tension field} of the section~$\phi$, and solutions of \(\tau^\VB(\phi)=0\) are called harmonic sections.
Note that \(\tau^\VB(\phi)\) coincides with the tension field \(\tau(\phi)\) if \(\dd\phi\) and \(\dd^\VB\phi\) coincide.
This happens, for example, in the case of the trivial action on \(N\) where \(\FB=M\times N\) and the connection is trivial.
Hence harmonic sections generalize harmonic maps to a gauged setting.

Recall that each section~$\phi\in\Gamma(\FB)$ corresponds uniquely to an equivariant map~$\tilde{\phi}\colon P\to N$, such that \(\iota\circ\left(\id_P, \tilde{\phi}\right) = \phi\circ \pi\).
In terms of~$\tilde{\phi}$ we have
\begin{align}
	\dd^\VB\phi(X)
	=\dd\iota_p\dd\tilde{\phi}\left(\tilde{X}-\Psi_p'\omega(\tilde{X})\right)
	=\dd\iota_p\left(\dd\tilde{\phi}(\tilde{X})+\dd\mu_{\tilde{\phi}(p)}\omega(\tilde{X})\right).
\end{align}
In addition, it is shown in~\cite{wood1990existence} that \(\phi\) is a harmonic section if and only if its corresponding \(G\)-equivariant map \(\tilde{\phi}\colon P\to N\) is harmonic with respect to the Kaluza--Klein metric~\(\fbm_P\) on~\(P\).
The Kaluza--Klein metric on \(P\) is given by
\begin{equation}
	\fbm_P(X, Y) = \left<\ver X, \ver Y\right>_\g + g_\pi\left(\hor X, \hor Y\right)
\end{equation}
where \(\left<\cdot, \cdot\right>_{\g}\) is an ad-invariant scalar product on \(VP=P\times\g\).

Analogous to the case without a gauge, the functional~\eqref{eq:energy of sections} is diffeomorphism invariant and in dimension two also conformally invariant.
Both lead to conservation laws by Noether's theorem.
Indeed, one can check that the energy-momentum tensor of is always divergence-free and, in dimension also two trace-free.

Despite the similarities to harmonic maps, the existence results do not immediately extend to harmonic sections because of the required equivariance properties.
For example, constant maps are trivially harmonic maps which generalize to the zero sections in the vector bundle case, but do not directly generalize to the case of bundles with nonlinear fibers.
It is shown in~\cite{wood1990existence} that the theory of the heat flow of harmonic maps can be used in certain cases to obtain harmonic sections:
If the fiber manifold~$(\fiber,\fm)$ has non-positive curvature and the fiber bundle~\(\FB\) allows for a \(C^1\)-section, then this section can be deformed via heat flow into a harmonic one.
For further exploration of harmonic section flow and applications, see e.g.~\cite{LE}.
In addition, when~$m=2$, the model possesses conformal invariance, and one can use the methods in~\cite{duzaar1998minimization,grotowski2004minimizing} to obtain harmonic sections in a given homotopy class.

\subsubsection{Local expressions}
For later convenience we derive the local expressions for the equations for harmonic sections.
The local representatives of the various geometric quantities will all be induced from a local section~$s\colon U\to \pi^{-1}(U)\subset P$ of the principal bundle~$P(M,G)$.

First, this local section~$s$ gives rise to a local trivialization of~$P$ over the domain of~$s$:
\begin{align}
	\chi^P_U \colon \pi^{-1}(U) \to   U\times G ,\qquad
	p \mapsto (\pi(p), \kappa(p))
\end{align}
where~$\kappa\colon \pi^{-1}(U)\to G$ is the \emph{structure group mapping} characterized by~$\Psi_{\kappa(p)}(s(\pi(p)))=p$.
It satisfies~$\kappa(s(x))=e\in G$, for any~$x\in U$, where~$e$ denotes the neutral element of~$G$.
Then the local form of~$\omega$ is given by
\begin{equation}
	A=s^*\omega\colon TU\to \g,
\end{equation}
that is,~$A$ is a~$\g$-valued one-form on~$U$.
Second,~$s$ also induces a local trivialization of the associated fiber bundle~$\FB$:
\begin{align}
	\chi^\FB_U \colon \rho^{-1}(U)  \to   U\times N \qquad
	[p, y] \mapsto \left(\rho([p, y])=\pi(p), \mu_{\kappa(p)}(y)\right).
\end{align}
The local representative of a section~$\phi\in\Gamma(\FB)$ is given by~$u\coloneqq\pr_2\circ\chi^\FB_U\circ\phi\colon U\to N$,
\begin{align}
	u(x)
	=\pr_2\circ \chi^\FB_U \left([s(x), \tilde{\phi}(s(x))]\right)=\mu_{\kappa(s(x))}\left(\tilde{\phi}(x)\right)
	=\tilde{\phi}(s(x))=\left(s^*\tilde{\phi}\right)(x).
\end{align}
That is,~$u=s^*\tilde{\phi}\colon U\to N$ and hence~$\chi^\FB_U\circ \phi(x)=(x, u(x))$ on~$U$.
Moreover, the tangent bundle of~$\FB$ is also locally trivialized:
\begin{align}
	T\left(\rho^{-1}(U)\right)\xrightarrow{\dd\chi^\FB_U} TU\times TN,
\end{align}
and the vertical differential of~$\phi$ takes the form
\begin{align}
	\pr_2\circ \dd\chi^\FB_U (\dd^\VB\phi_x(X))
	={}& \pr_2\circ \dd\chi^\FB_U \circ\dd\iota_p\left(\dd\tilde{\phi}_p(\hor\tilde{X})\right) \\
		& \qquad (\textnormal{note that} \pr_2\circ\chi^\FB_U\circ\iota_p(f)=\mu_{\kappa(p)}(f) ) \\
	={}& \dd\mu_{\kappa(p)}\left(\dd\tilde{\phi}_p(\hor\tilde{X})\right) \qquad \textnormal{(then use the $G$-equivariance)} \\
	={}& \dd\tilde{\phi}_{s(x)}\left( \dd\Psi_{{\kappa(p)}^{-1}} (\hor\tilde{X}) \right) \\
	={}& \dd\tilde{\phi}_{s(x)}\left(\hor\tilde{X}\right) \qquad \textnormal{(G-invariance of horizontal distributions)}
\end{align}
Here~$\tilde{X}$ is a lifting of~$X\in\Gamma(TU)$ to \(TP\).
In particular we could take~$\tilde{X}=s_*X$ and get
\begin{align}\label{eq:vertical differential in local trivialization}
	\pr_2\circ \dd\chi^\FB_U (\dd^\VB\phi_x(X))
	={}& \dd u(X)+ \dd\mu_{u(x)}(A(X))\equiv \dd_A u(X)\in \Gamma(u^* TN).
\end{align}
Furthermore, since~$\iota_p\colon (N,\fm)\to (\FB_x,\fbm)=(\rho^{-1}(x), \fbm)$ is an isometry for~$p=s(x)\in P$, we have, for a local orthonormal frame~$(e_\alpha)$ on~$U$ and writing~$\tilde{e}_\alpha=s_* e_\alpha$,
\begin{align}
	|\dd^\VB\phi|^2(x)
	={}& \sum_{\alpha} |\dd u(e_\alpha)+\dd\mu_{u(x)}(A(e_\alpha))|_{\fm}^2 =\sum_{\alpha}|\dd_A u(e_\alpha)|^2_{\fm}(x).
\end{align}
Therefore, locally we are considering the action
\begin{align}\label{eq:energy of sections-local}
	E(u;A)=\int_M |\dd_A u|^2_{g^\vee\otimes \fm}\dv_g=\int_{M} |\dd u+\dd\mu_u(A)|^2 \dv_g,
\end{align}
where~$g^\vee$ denotes the induced metric on the cotangent bundle.
Locally a variation~$(\phi_t)$ of~$\phi_0=\phi$ can be realized as~$\phi_t(x)=(x,u_t(x))\in U\times N$ where~$u_t\colon U\to N$ is a family of maps and the variational field is
\begin{equation}
	V=\dt{0}\phi_t= \dt{0} (\id, u_t) =(0,W),
\end{equation}
where~$W=\dt{0}u_t\in \Gamma(u^* TN)$.
A straightforward calculation shows
\begin{align}
	\dt{0}E(u_t;A)
	&=-2\int_M \left<W, \tau(u)+ \partial_1\partial_2\mu(A(e_\alpha),\dd u(e_\alpha))+ \dd\mu_u(\diverg(A))\right>\dv_g \\
	&\qquad -2\int_M \left<W, \partial_1\partial_2\mu\left(A(e_\alpha), \dd u(e_\alpha)\right)+\partial_1\partial_2\mu(A(e_\alpha),\dd\mu_u(A(e_\alpha)))\right>\dv_g.
\end{align}
Here the term~$\p_1\p_2\mu$ is defined as follows.
The differential of the group action \(\mu\colon G\times N\to N\) is given by \(\dd\mu\colon TG\times TN\to TN\) over~\(\mu\).
If we restrict it to the identity of \(G\), we obtain a bundle map \(\underline{\g}_{N}\oplus TN\to TN\) over~\(N\), still denoted by~$\dd\mu$, where \(\underline{\g}_{N}\) denotes the trivial bundle with fiber \(\g\) over \(N\).
Let now \(\mathfrak{a}\) be a section of the trivial bundle \(\underline{\g}_N\) and \(W\) a section of \(TN\).
Then \(\dd{\mu}(\mathfrak{a},W) = \dd{\mu}(\mathfrak{a},0) + W\) because \(\mu\) is the identity when restricted to \(e\in G\).
We will sometimes abbreviate~$\dd\mu(\mathfrak{a},0)$ as~$\dd\mu(\mathfrak{a})$ for simplicity, which can also be viewed as a partial tangent map (with fixed~$y\in N$).
Then we write
\begin{equation}
	\partial_1\partial_2\mu(\mathfrak{a},W)
	\equiv\nabla^N_W\dd\mu(\mathfrak{a},0)-\dd\mu(\nabla^{\g}_W \mathfrak{a},0)\in \Gamma(TN).
\end{equation}
where \(\nabla^\g_W \mathfrak{a}\) is the trivial covariant derivative on the trivial bundle \(\g\).
With respect to a basis~\(\epsilon_i\) of~$\g$ and \(\mathfrak{a}=\mathfrak{a}^i\epsilon_i\) we have \(\nabla^\g_W \mathfrak{a} = W(\mathfrak{a}^i)\epsilon_i\).
Notice that \(\partial_1\partial_2\mu(\mathfrak{a},W)\) is bilinear in \(\mathfrak{a}\) and \(W\) and can be seen as the off-diagonal part of the Hessian of \(\mu\).

Thus the Euler--Lagrange equations for the energy functional in terms of the local representative~$u$ reads
\begin{align}\label{eq:local equation}
	\tau(u)+2\partial_1\partial_2\mu(A(e_\alpha),\dd u(e_\alpha))+ \dd\mu_u(\diverg(A))+\partial_1\partial_2\mu(A(e_\alpha),\dd\mu_u(A(e_\alpha)))=0.
\end{align}
This is the local form of~$\tau^\VB(\phi)=0$.

\subsection{Coupling with Yang--Mills}
In this subsection, we recall some well-known geometric properties of the Yang--Mills action and study the dependence of the Dirichlet action on the principal connection.

The curvature of a principal connection \(\omega\) is the horizontal, equivariant \(\g\)-valued two-form~$\tilde{F}=D_\omega (\omega)$, satisfying
\begin{align}
	\tilde{F}&=\dd\omega+\frac{1}{2}[\omega,\omega], &
	D_\omega \tilde{F}&=0.
\end{align}
The horizontal, \(\ad\)-equivariant~$k$-forms on~$P$ can be reduced to~$k$-forms on~$M$ with values in~$\Ad(P)=P\times_{\Ad}\g$.
Equipping the compact Lie group~$G$ with a bi-invariant Riemannian metric~$\left<\cdot, \cdot\right>$, and hence~$\g$ with an~$\Ad$-invariant inner product~$\left<\cdot, \cdot\right>_\g$, we get a fiberwise Riemannian structure on~$\Ad(P)$, still denoted by~$\left<\cdot, \cdot\right>$ for simplicity.

In particular, the curvature can be identified with a section
\begin{equation}
	F=(e^\alpha\wedge e^\beta) \otimes F_{\alpha\beta}\in \Gamma\left((T^*M\wedge T^*M) \otimes_M\Ad(P)\right)
	\equiv \Omega^2(\Ad(P))
\end{equation}
with norm~$|F(x)|^2=\sum_{\alpha,\beta}\left< F_{\alpha\beta}(x),F_{\alpha\beta}(x)\right>$, where~$(e^\alpha)$ is a local orthonormal coframe on~$(M,g)$.
The Yang--Mills functional is
\begin{equation}
	\YM(\omega)=\int_M |F|^2 \dv_g.
\end{equation}

It is a fundamental point of gauge theory that the Yang--Mills action is invariant under gauge transformations, that is invariant under vertical automorphisms of the principal bundle.
We denote group of gauge transformations by~$\Gaug$.
Then for any $\varphi\in\Gaug$,
\begin{align}
	\varphi^*\omega &= \Ad_{\varphi^{-1}}(\omega), &
	\varphi^*(\tilde{F}) &= \Ad_{\varphi^{-1}}(\tilde{F}).
\end{align}
Hence~$|F|^2$ is gauge-invariant.

The variation formula for~$\YM(\omega)$ is standard, see, for example,~\cite{rudolph2017differential}:
The space of principal connections on~$P$ is an affine space~\(\Conn\) modeled on \(\Omega^1(\Ad(P))\).
More explicitly, fix an~$\omega\in\Conn$, then any other connection~\(\tilde{\omega}\) can be written as~\(\tilde{\omega}=\omega+\tilde{\zeta}\) for a unique horizontal, \(\ad\)-equivariant form \(\tilde{\zeta}\) which can be identified with \(\zeta\in \Omega^1(\Ad(P))\).
Let \(\omega_t = \omega + t\tilde{\zeta}\) be a variation of the connection \(\omega\) in the direction \(\tilde{\zeta}\) and denote the corresponding curvature tensor by~$F_t\in\Omega^2(\Ad(P))$.
Then
\begin{equation}\label{eq:varaition-YM}
	\dt{0}\int_M |F_t|^2\dv_g
	=2\int_M \left<D_\omega\zeta, F\right>
	=2\int_M \left< \zeta, D_\omega^* F\right>\dv_g,
\end{equation}
where~$D_\omega^*\colon \Omega^2(\Ad(P))\to \Omega^1(\Ad(P))$ is the adjoint of~\(D_\omega\colon \Omega^1(\Ad(P))\to \Omega^2(\Ad(P))\) with respect to the global~$L^2$ inner product on \(\Omega^*(\Ad(P))\).

Using the local section~$s$ as before and writing~$A=s^*\omega$, then~$F_A=s^*\tilde{F}$ satisfies
\begin{align}
F_A= \dd A+\frac{1}{2}[A,A], & & D_A F_A =\dd F_A +[A, F_A]=0.
\end{align}
Its codifferential is
\begin{equation}
	\begin{split}
		D_A^* F_A
		={}& \dd^*_A (\dd A+\frac{1}{2}[A,A])
		=\dd^*\dd A+\frac{1}{2}\dd^*[A,A]-A\llcorner\dd A-\frac{1}{2}A\llcorner [A,A].
	\end{split}
\end{equation}

We will now turn to the dependence of the Dirichlet \enquote{energy}~$E(\phi;\omega)$ on the gauge potential~$\omega$.
The Dirichlet action depends on the connection through the Kaluza--Klein metric and the vertical differential.

A gauge transformation \(\varphi\in\Gaug\) acts on a section~$\phi \in\Gamma(\FB)$ by
\begin{equation}
	\varphi(\phi)(x)= [\Psi_p(\varphi(\pi(p))), \tilde{\phi}(p)]
	=[p,\mu_{{\varphi(\pi(p))}^{-1}}(\tilde{\phi}(p))]
	\equiv [p,(\varphi^*\tilde{\phi})(p)].
\end{equation}
Moreover, the connection~$\varphi^*(\omega)$ induces a connection~$\varphi^*\sigma$ on~$T\FB$, given in the following way:
\begin{equation}
	\varphi^*\sigma (X_{\pi(p)}) = \dd\iota_y\left(\hor_{\varphi^*\omega}(\widetilde{X}_p)\right) \in T_{[p,y]}\FB.
\end{equation}
Thus the transformed vertical differential at~$x=\pi(p)$ is
\begin{align}
	\dd^\VB \varphi(\phi)(X_x)
	={}& \dd\iota_p\left(\dd{(\varphi^*\tilde{\phi})}_p(\tilde{X}_p)
		+ \dd\mu_{(\varphi^*\tilde{\phi})(p)}\varphi^*\omega(\tilde{X}_p)\right) \\
	={}& \dd\iota_p\left(\dd\mu_{{\varphi(x)}^{-1}}\dd\tilde{\phi}_p(\tilde{X}_p)+ \dd\mu_{{\varphi(x)}^{-1}} \dd\mu_{\tilde{\phi}(p)} \omega(\tilde{X}_p) \right).
\end{align}
Since~$\dd\iota_p$ and~$\dd\mu_{{\varphi(x)}^{-1}}$ both are isometries, we see that~$|\dd^\VB\phi|^2(x)= |\dd^\VB\varphi(\phi)|^2(x)$ and hence the energy term of the section is gauge invariant.

To derive the variation of the \(|\dd^\VB\phi|^2\) under a variation of the principal connection we pick a lift \((\tilde{e}_\alpha)\) of the \(g\)-orthonormal frame \((e_\alpha)\) to \(P\) and verify
\begin{align}
	|\dd^\VB\phi|_\fbm^2(x)
	&= \fbm_{\phi(x)}\left(\ver(\dd\phi(e_\alpha(x))),\ver(\dd\phi(e_\alpha(x)))\right)
	=\sum_{\alpha} |\dd\tilde{\phi}\left(\hor\tilde{e}_\alpha(p)\right)|_{\fm}^2,
\end{align}
where~$\pi(p)=x$.
Hence,
\begin{equation}
	\hor\tilde{e}_\alpha(p)
	=\tilde{e}_\alpha-\Psi'_p\left(\omega(\tilde{e}_\alpha(p))\right)
\end{equation}
is the only part depending on the connection~$\omega$.
Its derivative in the direction of \(\tilde{\zeta}\) is given by
\begin{align}
	\dt{0}\hor_t\dd\tilde{\phi}(e_\alpha(p))
	={}\dt{0} \left(\tilde{e}_\alpha-\Psi'_p\left(\omega_t(\tilde{e}_\alpha(p))\right)\right)
	={}\Psi'_p \left( \tilde{\zeta}(\tilde{e}_\alpha(p))\right)
	=\Psi'_p \left(\zeta(e_\alpha(x))\right)
\end{align}
and hence
\begin{align}
	\dt{0} |\dd\tilde{\phi}\left(\hor_t \tilde{e}_\alpha(p)\right)|^2_{\fm}
	={}&2 \left<\dd\tilde{\phi}\left(\Psi'_p\left(\zeta(e_\alpha(x))\right)\right), \dd\tilde{\phi}\left(\hor\tilde{e}_\alpha(p)\right)\right>_{\fm_{\tilde{\phi}(p)}}\\
	={}&2\left<\dd\iota_p\dd\mu_{\tilde{\phi}(p)}(\zeta(e_\alpha(x))), \dd\iota_p\dd\tilde{\phi}(\hor\tilde{e}_\alpha(p))\right>_{\fbm_{\phi(x)}}\\
	={}&2\left<\dd\bar{\mu}_{\phi(x)}(\zeta(e_\alpha(x))), \dd^\VB\phi(e_\alpha(x))\right>_{\fbm_{\phi(x)}},
\end{align}
where~$\dd\bar{\mu}_{\phi(x)}(\zeta(e_\alpha))$ is defined in the following way.
For a point~$z= [p, y]\in\FB$ consider the map
\begin{align}%
\label{eq:differential of mu from adjoint bundle}
	\dd\bar{\mu}_z\colon \Ad(P) &\to \VB_z, &
	[p,\zeta] &\mapsto [p,\dd\mu_{y}(\zeta)],
\end{align}
where~$\dd\mu_y$ is the differential of the evaluation map~$\mu_y\colon G\to N$ and~$\zeta\in\g$.
Denoting by~$\Ad(P)\times_M \FB$ the fiber product of~$\Ad(P)$ and~$\FB$ over~$M$, we have a well-defined map
\begin{equation}
	\dd\bar{\mu}\colon \Ad(P)\times_M \FB \to \VB.
\end{equation}
over the manifold~$\FB$. In particular, for a given section~$\phi\in\Gamma(\FB)$, there is an induced map
\begin{equation}\label{eq:adjoint action}
	\dd\bar{\mu}_\phi\colon \Gamma(\Ad(P))\to \Gamma(\phi^*\VB).
\end{equation}
Therefore we have
\begin{align}\label{eq:variation-vertical energy wrt A}
	\dt{0} \int_M |\dd^\VB\phi|_\fbm^2(x) \dv_g
	={}& 2\int_M \left<\dd\bar{\mu}_\phi(\zeta(e_\alpha)), \dd^\VB\phi(e_\alpha)\right>\dv_g \\
	={}& 2\int_M \left<\zeta, \dd\bar{\mu}_\phi^*(\dd^\VB\phi)\right>\dv_g,
\end{align}
where~$\dd\bar{\mu}^*_\phi$ denotes the formal~$L^2$-adjoint of~$\dd\bar{\mu}_\phi$ in~$\Hom(\Ad(P),\phi^*\VB)$.

Locally it is more explicit: variation of~\eqref{eq:energy of sections-local} with respect to the family~$(A_t=A+t\zeta)$ gives
\begin{equation}
	\dt{0}E(u;A_t)= 2\int_M \left<\dd\mu_u(\zeta), \dd_A u\right>\dv_g
	\equiv 2\int_M \left<\zeta, \dd\mu_u^*(\dd_A u)\right>\dv_g.
\end{equation}

\subsection{Coupling with Dirac}
Next we consider the coupling with the Dirac action, which is a technical part of the work.
From now on we assume that the base manifold~$(M,g)$ is spin and fix a spin structure.
Let~$S\to M$ be the associated spinor bundle with the Clifford map~$\gamma\colon TM\to \End(S)$ satisfying the Clifford relation
\begin{equation}
	\gamma(X)\gamma(Y)+\gamma(Y)\gamma(X)=-2g(X, Y), \qquad \forall X, Y\in\Gamma(TM).
\end{equation}
The Levi-Civita connection of~$(M,g)$ can be lifted to a connection on the spin principal bundle and thus induces a spin connection on~$S$.
We denote the corresponding covariant derivative by~$\nabla^s$.
The spin Dirac operator~$\pd s = \gamma(e_\alpha) \nabla^s_{e_\alpha}s$ is a first-order self-adjoint elliptic differential operator on \(S\).
Without loss of generality we assume that the spinor bundle~$S$ is always equipped with a~$\Spin(m)$ invariant metric~$g_s$.
For more about spin geometry we refer to~\cite{lawson1989spin, jost2008riemannian, ginoux2009dirac}.

It is important to note that the self-adjointness of \(\pd\) depends crucially on the minus sign in the Clifford relation; compare the discussion in~\cite{jost2018regularity}.
Without this minus sign, the Dirac operator would be anti-self-adjoint and the Dirac-term in the action below would vanish.
In the physics literature, the Clifford relation without minus sign is combined with anti-commuting spinors to to obtain a self-adjoint Dirac operator, see also~\cite{kessler2019super} and references therein.
The idea that anti-commuting variables can be avoided by using the minus sign in the Clifford relation in the study of actions coupling harmonic maps with spinors goes back to~\cite{chen2006dirac}.

In a sigma model, given a~$C^1$ section~$\phi\in\Gamma(\FB)$, we consider twisted spinorial fields along~$\phi$, that is, sections~$\psi\in\Gamma(S\otimes \phi^*\VB)$.
We still denote by $\gamma\colon TM\to \End(S\otimes\phi^*\VB)$ the Clifford map that arises from the Clifford map on \(S\) acting on the first factors.
The covariant derivative \(\LC\) on \(T\FB\) can be restricted to a covariant derivative on \(\VB\) denoted as \(\nabla^{\VB} = \ver \LC\).
Thus~$(S\otimes\phi^*\VB,\nabla^{S\otimes \phi^*\VB}, \gamma, g_s\otimes\phi^*\vbm)$ is a Dirac bundle, and the corresponding Dirac operator \(\D\psi = \gamma(e_\alpha) \nabla^{S\otimes\phi^*\VB}_{e_\alpha}\psi\) is again an essentially self-adjoint first-order elliptic operator.

The Dirac action of interest has the form
\begin{equation}
	\DA(\psi;\omega,\phi)=\int_M \left<\psi, \D\psi\right>_{g_s\otimes\phi^*\vbm}\dv_g.
\end{equation}

\subsubsection{Equations of motion}
Note that the spinor fields \(\psi\) depend on the section \(\phi\) and hence~\(\phi\) and \(\psi\) cannot be varied independently.
Let~$(\phi_t,\psi_t)$ be a variation family of~$(\phi=\phi_0,\psi=\psi_0)$ for~$t$ in a neighborhood of~$0$.
Noting that the spinor bundle does not change with~$t$, we have
\begin{align}
	\dt{0}\int_M & \left<\psi_t,\D^{\phi_t}\psi_t\right>\dv_g \\
	&= 2\int_M \left<\D\psi, \nabla^{S\otimes\phi_t^*\VB}_{\p_t}\psi_t\big|_{t=0}\right>\dv_g
		+\int_M \left<\psi, \gamma(e_\alpha)R^{\phi_t^*\VB}(\p_t, e_\alpha)\psi\big|_{t=0}\right>\dv_g.
\end{align}
The curvature term is tensorial in the variational field~$\phi_*(\p_t)$ and thus we define \(\VC(\phi, \psi)\in \Gamma\left({\left(\phi^*T\FB\right)}\right)\) by
\begin{equation}
	\left<\psi, \gamma(e_\alpha)R^{\phi_t^*\VB}(\p_t, e_\alpha)\psi\big|_{t=0}\right>
	\equiv\left<\phi_*(\p_t), \VC(\phi,\psi) \right>.
\end{equation}
Therefore the variation formula with respect to~$(\phi,\psi)$ is
\begin{equation}\label{eq:variation-no gauge}
	\dt{0}\DA(\phi_t,\psi_t;\omega)
	=\int_M 2\left<\D\psi, \nabla^{S\otimes\phi^*\VB}_{\p_t}\psi_t\big|_{t=0}\right>
		+\left<\VC(\phi,\psi), \phi_*(\p_t)\big|_{t=0}\right>\dv_g.
\end{equation}

\subsubsection{Local description}
Fix~\(x_0\in M\), \(y_0\in N\) and~\(p_0\in \pi^{-1}(x_0)\).
Let ${(x^\alpha)}_{\alpha=1,\dotsc,m}$ be normal coordinates in an open neighborhood \(U\) of~$x_0$.
We can choose a local section~\(s\colon U\to P\) such that the local representative~$A=s^*\omega$ satisfies \(A(x_0)=0\).
Let \(z^\nu\) be local coordinates on \(G\) around \(e\).
We denote the lift of \(x^\alpha\) and \(z^\nu\) to the product \(U\times G\) by \((\tilde{x}^\alpha, \tilde{z^\nu})\).

Let~\({(y_i)}_{i=1,\dotsc,n}\) be normal coordinates around \(y_0\).
The fiber bundle~\(\FB\) is locally around~\([p_0, y_0]\) a fiber product.
Denote the lift of the coordinates~\(x^\alpha\) and \(y^i\) to this product by \((\bar{x}^\alpha, \bar{y}^i)\), which forms a local coordinate system.
By construction, the local coordinate vector fields \(\partial_{y^i}\) are vertical, but the vector fields \(\partial_{x^\alpha}\) are not necessarily horizontal.

Indeed, noting that~$\bar{x}^\alpha\circ \iota_y=\tilde{x}^\alpha$ as local functions on~$P$,
\begin{align}
	\hor\left(\frac{\p}{\p \bar{x}^\alpha}\right)
	=& \dd\iota_y\left(\hor \frac{\p}{\p\tilde{x}^\alpha}\right)
	=\frac{\p}{\p\bar{x}^\alpha}
		+\dd\iota_p\circ \dd\mu_{y}\circ \omega\left(\frac{\p}{\p \tilde{x}^\alpha}\right),\\
	\ver\left(\frac{\p}{\p\bar{x}^\alpha}\right)
	=&-\dd\iota_p\circ \dd\mu_{y}\circ \omega\left(\frac{\p}{\p \tilde{x}^\alpha}\right)
	=-\dd\bar{\mu}_z\circ A\left(\frac{\p}{\p x^\alpha}\right).
\end{align}
Thus
\begin{align}
	\fbm_{\alpha\beta}
	&= \fbm\left(\frac{\p}{\p \bar{x}^\alpha}, \frac{\p}{\p \bar{x}^\beta}\right)
		= g\left(\frac{\p}{\p x^\alpha},\frac{\p}{\p x^\beta}\right)
		+ \vbm\left(\ver\left(\frac{\p}{\p \bar{x}^\alpha}\right), \ver\left(\frac{\p}{\p \bar{x}^\beta}\right)\right) \\
	&=g_{\alpha\beta}+ \vbm\left(\dd\bar{\mu}_z A\left(\frac{\p}{\p x^\alpha}\right), \dd\bar{\mu}_z A \left(\frac{\p}{\p x^\beta}\right)\right)
	\equiv g_{\alpha\beta}+ \vbm_{\alpha\beta}, \\
	\fbm_{ij}
	&=\fbm\left(\frac{\p}{\p\bar{y}^i}, \frac{\p}{\p \bar{y}^j}\right) = \fm_{ij}, \\
	\fbm_{\alpha i}
	&= \fbm\left(\frac{\p}{\p\bar{x}^\alpha}, \frac{\p}{\p\bar{y}^i}\right)
		=\vbm\left(\ver\left(\frac{\p}{\p \bar{x}^\alpha}\right), \frac{\p}{\p\bar{y}^i}\right)
		=-\vbm\left(\dd\bar{\mu}_z A \left(\frac{\p}{\p x^\alpha}\right),\frac{\p}{\p \bar{y}^i}\right).
\end{align}
Notice that we use Greek indices for coordinates of the base and Latin fiber indices.
At the point~$[p_0, y_0]$,
\begin{align}%
\label{eq:fbm at normal coord}
	\fbm_{\alpha\beta}=\delta_{\alpha\beta}, & & \fbm_{ij}=\delta_{ij}, & & \fbm_{\alpha i}=0,
\end{align}
and all the Christoffel symbols of \(\LC\) vanish at this given point.

Any spinor field~$\psi$ along the section~$\phi$ can be expressed as
\begin{equation}
	\psi=\psi^i\otimes \phi^*\left(\frac{\p}{\p\bar{y}^i}\right).
\end{equation}
The vertical connection acts on such spinors in the following way: for any~$X\in\Gamma(TM)$,
\begin{equation}
	\nabla^{S\otimes\phi^*\VB}_X\psi
	= \nabla^s_X \psi^i\otimes \phi^*\left(\frac{\p}{\p \bar{y}^i}\right)
		+\psi^i\otimes \nabla^{\phi^*\VB}_X \phi^*\left(\frac{\p}{\p \bar{y}^i}\right),
\end{equation}
where
\begin{equation}
	\nabla^{\phi^*\VB}_X\phi^*\left(\frac{\p}{\p\bar{y}^i}\right)
	=\ver\phi^*\left(\LC_{\dd\phi(X)}\frac{\p}{\p\bar{y}^i}\right).
\end{equation}
Writing~$\dd\phi(X)=X(\phi^\beta)\p_{\bar{x}^\beta}+ X(\phi^j)\p_{\bar{y}^j}$ and since the fibers are totally geodesic, we have
\begin{align}%
\label{eq:vertical part of a general connection term}
	\nabla^{\phi^*\VB}_X\phi^*\left(\frac{\p}{\p\bar{y}^i}\right)
	&= \ver\left(X(\phi^\beta)\LC_{\p_{\bar{x}^\beta}} \p_{\bar{y}^i} + X(\phi^j)\LC_{\p_{\bar{y}^j}} \p_{\bar{y}^i} \right) \\
	&=-X(\phi^\beta)\Gamma_{\beta i}^\eta \dd\bar{\mu}_{\phi(x)} A\left(\frac{\p}{\p x^\eta}\right)
		+X(\phi^\beta)\Gamma_{\beta i}^k \frac{\p}{\p\bar{y}^k}
		+X(\phi^j)\Gamma_{ji}^k \frac{\p}{\p \bar{y}^k}.
\end{align}
The associated Dirac operator~$\D$ on~$S\otimes\phi^*\VB$ is defined in the canonical way: taking a local orthonormal basis~$(e_\alpha)$ on~$M$, for any spinor~$\psi$ along the section~$\phi$,
\begin{align}
	\D\psi
	&=\gamma(e_\alpha)\nabla^{S\otimes\phi^*\VB}_{e_\alpha}\psi
	=\pd\psi^i\otimes \phi^*\left(\frac{\p}{\p\bar{y}^i}\right)
		+\gamma(e_\alpha)\psi^i\otimes \ver \phi^*\left(\LC_{\dd\phi(e_\alpha)}\frac{\p}{\p\bar{y}^i}\right).
\end{align}

\subsubsection{Dependence on the gauge potential}
The Dirac action is gauge invariant because the Lie group~$G$ does not act on the pure spinor bundle~$S$ while it acts on~$(N,h)$ via isometries.
A local argument was also suggested in~\cite{isobe2010regularity}.
We need to consider the variation with respect to the gauge potential~$\omega$.
As before we consider~$\omega_t=\omega+t\tilde{\zeta}$.
Note that the Kaluza--Klein metric~$\fbm$ depends on~$\omega_t$ via the Ehresmann connections~$\sigma_t$, namely the horizontal and vertical projections:
\begin{equation}
	\fbm_t(X, Y)
	= g_\rho(\hor_t X, \hor_t Y) + \vbm(\ver_t X, \ver_t Y),
\end{equation}
while the vertical metric~$\vbm$ does not.
Hence
\begin{align}%
\label{eq:variation of Dpsi with gauge potential}
	\frac{\dd}{\dd t} \D_t\psi
	={}& \gamma(e_\alpha)\psi^i\otimes \frac{\dd}{\dd t} \phi^*\left(\nabla^{t,\VB}_{\phi_*(e_\alpha)}\frac{\p}{\p\bar{y}^i}\right).
\end{align}
Thus, the problem is reduced to analyze the dependence of \(\nabla^\VB\) on the connection \(\omega\).

Consider the coordinates~$(\bar{x}^\alpha,\bar{y}^i)$ of \(\FB\) around~\([p_0, y_0]\).
For a general~$t\neq 0$, the local vectors~${\{\p/\p {\bar{y}^i}\}}_{1\le i\le n}$ stay orthonormal at~$[p_0, y_0]$ and vertical, while the vectors~$\{\p/\p\bar{x}^\alpha\}$ are in general neither horizontal nor orthonormal.
Hence,
\begin{align}
	\dt{0}\fbm_{ij}
	&= \dt{0}\vbm_{ij}=0, \\
	\dt{0}\fbm_{\alpha\beta}
	&= \vbm\left(\dt{0}\ver_t(\frac{\p}{\p\bar{x}^\alpha}), \ver(\frac{\p}{\p\bar{x}^\beta})\right)
		+\vbm\left(\ver(\frac{\p}{\p\bar{x}^\alpha}),\dt{0}\ver_t(\frac{\p}{\p\bar{x}^\beta})\right), \\
	\dt{0}\fbm_{\alpha i}
	&= \vbm\left(\dt{0}\ver_t(\frac{\p}{\p\bar{x}^\alpha}), \ver(\frac{\p}{\p\bar{y}^i})\right).
\end{align}
The~$t$-derivative of the vertical parts is given by
\begin{align}
	\dt{0}\ver_t(\frac{\p}{\p\bar{x}^\alpha})
	={}& \dt{0}-\dd\iota_p \dd\mu_y \omega_t\left(\frac{\p}{\p\tilde{x}^\alpha}\right)
	= -\dd\bar{\mu}_z \circ \zeta\left(\frac{\p}{\p x^\alpha}\right).
\end{align}
Substituting this into the above formulas, we get
\begin{align}
	\dt{0}\fbm_{\alpha\beta}
	={}& \vbm\left( \dd\bar{\mu}_z \zeta\left(\frac{\p}{\p x^\alpha} \right), \dd\bar{\mu}_z A \left(\frac{\p}{\p x^\beta}\right)\right)
		+\vbm\left( \dd\bar{\mu}_z A\left(\frac{\p}{\p x^\alpha} \right), \dd\bar{\mu}_z \zeta\left(\frac{\p}{\p x^\beta}\right)\right) \\
	={}& (\bar{\mu}_z^* \vbm) \left(\zeta\left(\frac{\p}{\p x^\alpha} \right), A\left(\frac{\p}{\p x^\beta}\right)\right)
		+(\bar{\mu}_z^* \vbm) \left(A\left(\frac{\p}{\p x^\alpha} \right), \zeta\left(\frac{\p}{\p x^\beta}\right)\right), \\
	\dt{0}\fbm_{\alpha i}
	={}& -\vbm\left(\dd\bar{\mu}_z \zeta\left(\frac{\p}{\p x^\alpha}\right), \frac{\p}{\p \bar{y}^i}\right).
\end{align}
Now we continue to compute~\eqref{eq:variation of Dpsi with gauge potential}.
The points under consideration are~$\phi(x)=[p,y]\in\FB$ and~$y=\tilde{\phi}(p)\in \fiber$.
Write
\begin{equation}
	\phi_*(e_\alpha)= \phi_\alpha^\beta \frac{\p}{\p \bar{x}^\beta}+\phi_\alpha^j\frac{\p}{\p \bar{y}^j},
\end{equation}
and denote the Christoffel symbols of~$\fbm(t)$ by~$\Gamma(t)$, and~$A_t =A+t\zeta$, then
\begin{equation}
	\ver_t\nabla^{\fbm(t)}_{\dd\phi(e_\alpha)}\frac{\p}{\p\bar{y}^i}
	=\phi_\alpha^\beta \Gamma_{\beta i}^\eta (t)\dd\bar{\mu}_{\phi(x)} A_t\left(\frac{\p}{\p\bar{x}^\eta}\right)
	+\phi_\alpha^\beta \Gamma_{\beta i}^k(t)\frac{\p}{\p \bar{y}^k}
	+ \phi_\alpha^j \Gamma_{ji}^k (t) \frac{\p}{\p \bar{y}^k}.
\end{equation}
Note that~$\dt{0}\Gamma_{ji}^k=0$ since the vertical part does not involve the connection, while for~$\Gamma^k_{\beta i}$ at~$\phi(x_0)\in \FB$,
\begin{align}
	\dt{0}\Gamma_{\beta i}^k
	={}&\frac{1}{2}\left[\frac{\p}{\p\bar{y}^i}\dt{0}\fbm_{k\beta}-\frac{\p}{\p\bar{y}^k}\dt{0}\fbm_{\beta i}\right] \\
	={}&\frac{1}{2}\left[\bar{h}\left(\partial_1\partial_2\bar{\mu}_{\phi(x)}(\zeta(\frac{\p}{\p x^\beta}),\frac{\p}{\p\bar{y}^i}),\frac{\p}{\p\bar{y}^k}\right)
		-\bar{h}\left(\partial_1\partial_2\bar{\mu}_{\phi(x)}(\zeta(\frac{\p}{\p x^\beta}),\frac{\p}{\p\bar{y}^k}),\frac{\p}{\p\bar{y}^i}\right)\right]\\
	={}&\bar{h}\left(\partial_1\partial_2\bar{\mu}_{\phi(x)}(\zeta(\frac{\p}{\p x^\beta}),\frac{\p}{\p\bar{y}^i}),\frac{\p}{\p\bar{y}^k}\right),
\end{align}
where in the last step we used the skew-symmetry of~$\p_1\p_2\mu$:
\begin{equation}
	\left<\p_1\p_2\mu(\mathfrak{a},Y),\;Z\right>_\fm
	+\left<Y,\;\p_1\p_2\mu(\mathfrak{a},Z)\right>=0,
\qquad \forall \mathfrak{a}\in\mathfrak{g}, \;\;
	\forall Y,Z \in\Gamma(TN).
\end{equation}
We thus get
\begin{align}
	\dt{0}\nabla^{t,\VB}_{\phi_*(e_\alpha)}\frac{\p}{\p\bar{y}^i}
	={}&\phi^\beta_\alpha \bar{h}\left(\partial_1\partial_2\bar{\mu}_{\phi(x)}(\zeta(\frac{\p}{\p x^\beta}),\frac{\p}{\p\bar{y}^i}),\frac{\p}{\p\bar{y}^k}\right) \frac{\p}{\p\bar{y}^k} \\
	={}& \partial_1\partial_2\bar{\mu}_{\phi(x)}\left(\zeta(\frac{\p}{\p x^\alpha}),\frac{\p}{\p\bar{y}^i}\right)
	\equiv  \left<\zeta, e_\alpha\otimes \partial_1\partial_2\bar{\mu}_{\phi(x)}\left(\frac{\p}{\p\bar{y}^i}\right) \right>.
\end{align}
It follows that
\begin{align}
	\dt{0}\D_t\psi=\gamma(e_\alpha)\psi^i\otimes \partial_1\partial_2\bar{\mu}_{\phi(x)}\left(\zeta(e_\alpha),\frac{\p}{\p\bar{y}^i}\right),
\end{align}
and
\begin{align}\label{eq:variation-Dirac with gauge}
	\dt{0}\left<\psi,\D_t\psi\right>
	=\left<\psi^j,\gamma(e_\alpha)\psi^i\right>\cdot
		\bar{h}\left(\partial_1\partial_2\bar{\mu}_{\phi(x)}(\zeta(e_\alpha),\frac{\p}{\p\bar{y}^i}),\frac{\p}{\p\bar{y}^j}\right)
	\equiv\left<\zeta, \Qpsi(\phi,\psi)\right>.
\end{align}
Note that both factors in the middle are anti-symmetric in~$i$ and~$j$.

\subsection{The coupled action}%
\label{subsect: coupled action}
In the remainder of this article we will be concerned with the model with the action
\begin{equation}\label{eq:coupled action}
	\YMHD(\omega,\phi,\psi)
	= \YM(\omega) + E(\phi; \omega) + \DA(\psi; \omega)
	=\int_M |F|^2+|\dd^\VB\phi|^2+\left<\psi,\D\psi\right>\dv_g.
\end{equation}
This \emph{Yang--Mills--Higgs--Dirac action} might also be considered as a gauged version of Dirac-harmonic maps.
We have already proven that the Yang--Mills--Higgs--Dirac action is gauge invariant and that, thanks to~\eqref{eq:variation-vertical harmpnic maps},\eqref{eq:varaition-YM},\eqref{eq:variation-no gauge} and~\eqref{eq:variation of Dpsi with gauge potential},\eqref{eq:variation-Dirac with gauge}, its total variation formula has the form
\begin{align}
	\delta\YMHD
	=\int_M & \left<\zeta,D_\omega^*F\right>-2\left<\tau^\VB(\phi),\delta\phi\right>+\left<\zeta,\dd\bar{\mu}^*_\phi(\dd^\VB\phi)\right> \\
		& +2\left<\D\psi, \delta\psi\right>+\left<\VC(\phi,\psi),\delta\phi\right>
			+\left<\zeta, \Qpsi(\phi,\psi) \right>\dv_g,
\end{align}
where~$\tilde{\zeta}=\delta\omega$ as before.
Hence, we have shown:
\ELEquations{}
This is a coupled system, with the equation for~$\phi$ and~$\psi$ being elliptic.
The equation for~$\omega$ is actually also (locally) elliptic up to a local gauge, as explained in Section~\ref{Sec:Regularity} below.

Let us now explain the scaling behavior of the different terms in the action.
For convenience we take~$U$ to be the unit disk~$B_1(0)$ with Euclidean metric and assume the bundles are trivialized there.
For~$r>0$, denote the dilation
\begin{equation}
	\theta_r\colon B_1(0)\to B_r(0), \qquad x\mapsto rx.
\end{equation}
With respect to the Euclidean metric~$g_0$ on both sides,~$\theta_r$ is conformal:~$\theta_r^* g_0= r^2 g_0$.
It is not hard to see that the fields involved are scaled in the following way:
\begin{lemma}\label{lemma:scaling behaviors}
	Consider the trivial bundle~$P_r=B_r(0)\times G \to B_r(0)$ with connection form~$A$ and let~$u \colon B_r(0)\to \fiber$ be a section of the (associated) fiber bundle.
	Let~$A_r(x)\equiv rA(rx)$ be the connection form on~$B_1(0)$ for the pullback bundle~$\theta_r^* (P_r)$,
	while~$u_r(x)\coloneqq \theta_r^* u(x)= u(rx)$ and~$\psi_r(x)=r^{\frac{m-1}{2}}\psi(rx) \in \Gamma(S\otimes u_r^* TN \to B_r(0))$.
	Then
	\begin{align}
		\int_{B_1(0)} |F(A_r)|^2 \dx
		&=r^{4-m} \int_{B_r(0)} |F(A)|^2 \dx, \\
		\int_{B_1(0)} |\dd_{A_r} (\theta_r^* u)|^2 \dx
		&=r^{2-m}\int_{B_r(0)} |\dd_A u|^2 \dx, \\
		\int_{B_1(0)} \left<\D^{u_r}\psi_r,\psi_r\right>\dx
		&= \int_{B_r(0)} \left<\D\psi, \psi \right>\dx, \\
		\int_{B_1(0)} |\psi_r|^{\frac{2m}{m-1}}\dx
		&= \int_{B_r(0)} |\psi|^{\frac{2m}{m-1}} \dx.
	\end{align}
\end{lemma}
This implies that, for~$r\in (0,1)$, the Dirac term stays rescaling invariant, as well as the~$L^{\frac{2m}{m-1}}$ norm of the spinors.
We see that dimension two is critical for the Dirichlet energy part of the action in the general case of nonlinear fibers and the Yang--Mills energy part is conformally invariant in dimension four. In the remainder of the article we will investigate the regularity of weak solutions and their blow-up behavior in the lowest critical dimension, that is, dimension two.

While we focus for simplicity on the Yang--Mills--Higgs--Dirac action, several extensions have been considered in the literature:
\begin{itemize}
	\item
		Instead of~\(\DA\) one might consider a massive Dirac action given by
		\begin{equation}
			\int_M \left<\psi, \D\psi\right> - \kappa|\psi|^2\dv_g,
		\end{equation}
		see, for example,~\cite{parker1982gauge}.
		Here the parameter~$\kappa\in\R$ is interpreted as the mass of the spinors in physics.
		In this case, the Dirac equation is \(\D\psi=\kappa\psi\).
		However, the mass term behaves badly under scaling and is dropped in our analysis.
	\item
		In addition to the Yang--Mills--Higgs--Dirac action one might consider a curvature term for the spinors~\(\psi=\psi^i\otimes\phi^*\partial_{y^i}\):
		\begin{equation}
			\frac16\int_M g_s(\psi^i, \psi^k)g_s(\psi^j, \psi^l)\fbm(R^\FB(\partial_{y^i}, \partial_{y^j})\partial_{y^k}, \partial_{y^l}) \dv_g.
		\end{equation}
		The derivation of the additional terms in the equations of motion is straightforward, compare also~\cite{Branding2015Curvature}.
		This is relevant in physics when the fibers are Kähler or Calabi--Yau manifolds.
	\item
		More generally, an additional potential term~$\potential$ is often needed in physics,
		\begin{equation}
			\Action_W(\omega,\phi,\psi)
			=\int_M |F(\omega)|^2 + |\dd^\VB\phi|^2 + \potential(\phi)
				+ \left<\psi,\D\psi\right>\dv_g,
		\end{equation}
		where~$\potential\colon \FB\to\R$ stands for a~$G$-invariant function.
		For example, when the fiber is a vector space a polynomial potential is usually used and when the fiber is symplectic, the momentum map is used, see e.g.~\cite{riera1999thesis, FH, WZ, riera2009compactification, song2011critical, song2016convergence, ASZ}.
		We do not include this potential term since it does not affect our analysis too much in dimension two, as long as the integrability of the potential is guaranteed and certain abstract growth conditions are posed.
		More generally the potential term could also depend on the spinorial field, and it is then helpful to obtain minimax solutions, see~\cite{isobe2011existence, isobe2011nonlinear, yang2017solutions}.
	\item
		Instead of the Levi-Civita connection \(\LC\) on \(\FB\) one might consider more general metric connection, allowing for torsion, compare also~\cite{branding2016dirac}.
	\item
		In~\cite[Chapter~6]{deligne1999supersolutions}, a fully supersymmetric variant of the Yang--Mills--Higgs--Dirac action is given, which has motivated our study here.
		The fully supersymmetric theory requires an additional twisted spinor \(\lambda\in \Gamma(S^*\otimes \ad P)\) as a superpartner of the connection.
		The action for \(\lambda\) is also the Dirac action together with lower order terms coupling to~$\phi$ and~$\psi$.
		In case the equation for the additional spinorial field is subcritical, the analysis could be carried out by extending the methods here.
		Notice, however, that we cannot expect full supersymmetry in our model, even when extended by \(\lambda\).
		The reason is that supersymmetry requires anti-commuting variables which we are avoiding for the sake of analysis.
	\item
		A BRST type symmetry was indicated in~\cite{witten1988topological} for a similar model, which can be generalized to the gauged setting, if one uses anti-commuting variables.
		Then one has to take into considerations the curvature terms and suitable potentials in the action functional.
		The related constructions will follow in a straightforward manner from our treatment of the main terms.
\end{itemize}

\section{Regularity of weak solutions}%
\label{Sec:Regularity}
In the remaining part we focus on the critical dimension, that is, we consider a Riemann surface~$M$ as domain.
First we show regularity of critical points up to a gauge transformation.

In this case the Yang--Mills--Higgs--Dirac action is naturally defined on the space
\begin{equation}
	\dom(\YMHD)\coloneqq \left\{(\omega,\phi,\psi)\mid \omega\in\Conn^{1,2}, \phi\in W^{1,2}(\Gamma(\FB)), \psi\in W^{1,\frac{4}{3}}(\Gamma(S\otimes \phi^*\VB))\right\}.
\end{equation}

\begin{defn}
	A triple~$(\omega,\phi,\psi)\in\dom(\YMHD)$ is called a weak solution of the system~\eqref{eq:EL-global} if it satisfies the system~\eqref{eq:EL-global} in the sense of distributions.
	More precisely, for any smooth triple~$(\zeta, V, \eta)$ with~$\zeta\in \Gamma(\Ad(P))$,~$V\in \Gamma(\phi^*\VB)$, and~$\eta\in \Gamma(S\otimes\phi^*\VB)$, it holds that
	\begin{multline}
		\int_M \left<D\zeta, F(\omega)\right>
		+2\left<\nabla^{\phi^*\VB}_{e_\alpha}V, \dd^\VB\phi(e_\alpha) \right>
		+\left<\dd\bar{\mu}(\zeta),\dd^\VB\phi\right>\dv_g \\
		+\int_M 2\left<\psi,\D^\phi\eta\right>
		+\left<\psi,\gamma(e_\alpha)R^{\phi^*\VB}(V,\dd^\VB\phi(e_\alpha))\psi\right>\dv_g \\
		+\int_M \left<\psi^i,\gamma(e_\alpha)\psi^j\right> \left<\partial_1\partial_2\bar{\mu}\left(\zeta(e_\alpha),\p_{y^i}\right), \p_{y^j}\right> \dv_g =0.
	\end{multline}
\end{defn}
We can also require the test functions~$(\zeta, V,\eta)$ to be of regularity~$C^1$.
The aim of this section is to prove the following
\RegularityOfWeakSolution

Here, the gauge transformation is needed because the Euler--Lagrange equation for the connection fails to be elliptic.
However, thanks to a result by K. Uhlenbeck\cite{uhlenbeck1982connections}, we can choose a gauge transformation making the equation locally elliptic.
Indeed, let~$\Conn^{k,p}$ be the space of~$W^{k,p}$-connections, and~$\Gaug^{k+1, p}$ the space of~$W^{k+1, p}$ gauges.
Then we know that~$\Gaug^{k+1,p}$ acts on~$\Conn^{k,p}$.
\begin{prop}[{\cite[Lemma 1.2]{uhlenbeck1982connections}}]%
\label{prop:regularity of gauge}
	Let~$(k+1)p>m=\dim M$.
	Then
	\begin{enumerate}
		\item The gauge group~$\Gaug^{k+1,p}$ is a smooth Lie group.
		\item The induced map
					\begin{equation}
					\Gaug^{k+1,p}\times \Conn^{k,p} \to \Conn^{k,p}, \quad
					(\varphi, \omega)\mapsto \varphi^*\omega
					\end{equation}
					is smooth.
		\item If~$\omega$ and~$\varphi^*\omega$ both are in~$\Conn^{k,p}$, then the gauge transformation~$\varphi$ has regularity~$W^{k+1,p}$, i.e.~$\varphi\in \Gaug^{k+1,p}$.
	\end{enumerate}
\end{prop}

\begin{theorem}[{\cite[Theorem 2.1]{uhlenbeck1982connections},~\cite[Theorem 6.1]{wehrheim2004Uhlenbeck}}]%
\label{thm:Uhlenbeck Coulomb gauge}
	Let~$p\in (\frac{m}{2}, m]$ and~$G$ be compact.
	Consider a connection~$\omega$ on the bundle~$ B_1(0)\times G$ with local representative $\tilde{A}$.
	Then there exist~$\kappa=\kappa(m)>0$ and~$c=c(m)>0$ such that if~$\|F(\tilde{A})\|_{L^{m/2}(B_1)}\le \kappa$, then~$\tilde{A}$ is gauge equivalent to a local connection form~$A$ such that
	\begin{enumerate}
		\item $\dd^* A=0$;
		\item $(x\cdot A)=0$ on~$\p B_1(0)$;
		\item $\|A\|_{W^{1,m/2}} \le c(m)\|F(\tilde{A})\|_{L^{m/2}}$;
		\item $\|A\|_{W^{1,p}} \le c(m)\|F(\tilde{A})\|_{L^{p}}$.
	\end{enumerate}
\end{theorem}
The gauge transformation in the above theorem is usually referred to as a \emph{Coulomb gauge}.
We remark that in~\cite{uhlenbeck1982connections} the theorem was stated with~$p\in(\frac{m}{2}, m)$, while it actually works for~$p\ge \frac{m}{2}$, see~\cite[Chapter 6]{wehrheim2004Uhlenbeck}.

The strategy to prove~\ref{thm:regularity of weak solution} is similar to the one for harmonic maps, but in addition, we need to glue the local gauges together to get a good Coulomb global gauge.
Note that~$m=2$ is a subcritical dimension for the Yang--Mills part, thus we can easily improve the regularity for the connection, at least locally.

\begin{proof}[Proof of Theorem~\ref{thm:regularity of weak solution}]
We first deal with the local regularity in a suitable gauge and then glue the local gauge to obtain the global smoothness.

\noindent \textbf{Step 1. Local regularity.}
	Let us take a local geodesic ball, say~$B_1(0)$, on which the fiber bundle is trivialized:~$\FB|_{B_1(0)}\cong B_1(0)\times N$.
	Embed~$N$ isometrically into some Euclidean space~$\R^K$, with second fundamental form~$\II$.
	Then~$\dd^\VB\phi$ is represented by
	\begin{equation}
			\dd_A u= \dd u + \dd\mu_u(A) \in \Gamma(T^* M \otimes u^*TN \to B_1(0)).
	\end{equation}
	The spinor along the section~$\phi$ is now locally a spinor along the map~$u\colon B_1(0)\to N$, and with respect to a local (normal) coordinate system~$(y^i)$ the spinorial field takes the form
	\begin{equation}
			\psi=\psi^i\otimes u^*\left(\p_{y^i}\right)\in \Gamma(S\otimes u^* TN).
	\end{equation}
	A basis for~$\g$ is denoted by~$(\epsilon_a)$, $1\le a \le \dim G$, with dual basis~$\epsilon^a$.
	Then~$(A,u,\psi)$ satisfies the equations in~\eqref{eq:EL-global} weakly on~$B_1(0)$:
	\begin{align}\label{eq:EL-local}
		\dd^*\dd A
		={}&-\frac{1}{2}\dd^*[A,A]+A\llcorner\dd A
				+\frac{1}{2}A\llcorner [A,A]
				-{(\dd\mu_u)}^t\left(\dd u+ \dd\mu_u(A)\right)\\
		& -\left<\psi^j,\gamma(e_\alpha)\psi^i\right>
				\left<\partial_1\partial_2\bar{\mu}(\epsilon_a,\p_{y^i}),\p_{y^j}\right>
				\epsilon^a\otimes e_\alpha, \\
		\Delta u
		={}& \tr\II(u)(\dd u,\dd u)
				-2\tr\partial_1\partial_2\mu(A,\dd u)
				-\dd\mu_u(\diverg A)
				-\tr\partial_1\partial_2\mu_u \left(A,\dd\mu(A)\right) \\
		& +\frac{1}{2}\left<\psi^i,\gamma(e_\alpha)\psi^j\right>
			\left<\p_{y^i}, R\left(\p_{y^k}, \dd u(e_\alpha)
			+\dd\mu_u(A(e_\alpha))\right)\p_{y^j}\right> h^{kl}(u) u^*(\p_{y^{l}}),\\
		\pd\psi^i
		={}&
		\left\{- \Gamma^\eta_{\alpha k}(u){\left(\dd\mu_u A(\p_{x^\eta})\right)}^i+ \Gamma^i_{\alpha k}(u)+\Gamma^i_{jk}(u) u^j_\alpha\right\}\gamma(e_\alpha)\psi^k
	\end{align}
	Thanks to Theorem~\ref{thm:Uhlenbeck Coulomb gauge}, by applying a Coulomb gauge if necessary, we may assume from the beginning that the local trivialization is chosen such that~$\dd^* A=0$.
	This is possible since along such a rescaling the energy for~$u$ is preserved while the energy for~$A$ is decreased.
	Therefore,~$A$ satisfies a local Laplace equation which is elliptic.

	Now, the equation for~$A$ and the regularity assumptions on the weak solutions, implies that~$A\in W^{2,p}_{loc} (B_1(0))$ for any~$p\in [1,2)$.
	In particular,~$A\in W^{1,q}_{loc}(B_1(0))$ for any~$q\in[1,+\infty)$.

	Then we turn to the spinor field~$\psi$.
	Applying~\cite[Lemma 6.1]{jost2018regularity} to this equation we get that~$\psi\in L^p_{loc}(B_1(0))$ for any~$p\in [1,+\infty)$.
	Then the regularity theory for the Dirac operator~$\pd$ implies~$\psi\in W^{1,2}_{loc}(B_1(0))$.

	The equation for~$u$ can be rewritten in the form
	\begin{equation}
	-\Delta u= \Omega\cdot \nabla u + f
	\end{equation}
	where~$\Omega\in L^2$ is an antisymmetric matrix and~$f=f(A,\psi, u)\in L^p_{loc}(B_1(0))$ for any~$p\in [1,2)$.
	Thanks to the regularity theory developed in~\cite{riviere2007conservation, riviere_struwe, riviere2010conformally, sharp2013decay}, we conclude that~$u\in W^{2,p}_{loc}(B_1(0))$ for any~$p\in [1,2)$.

	Now the situation is subcritical for all the fields, and a bootstrap argument then implies that they are actually in~$C^\infty(B_{1/2}(0))$.

\noindent \textbf{Step 2. Gluing.}
	We may suppose that there is a finite open cover~${\{U_\alpha\}}_{1\le \alpha \le l}$ such that each~$U_\alpha$ is a geodesic ball, and on each~$U_\alpha$ there exists a Coulomb gauge~$\varphi_\alpha$ such that the triple~$(\varphi_\alpha^*\omega, \varphi_\alpha^*\phi,\varphi_\alpha^*\psi)$ is smooth on~$U_\alpha$.

	On an intersection~$U_\alpha\cap U_\beta$, the two connections~$\varphi_\alpha^*\omega$ and~$\varphi_\beta^*\omega$ are both smooth.
	Therefore by Proposition~\ref{prop:regularity of gauge} the gauge~$\varphi_\alpha^{-1}\circ \varphi_\beta$ is smooth.
	Moreover, by precomposing with a smooth gauge if necessary, we may assume that both~$\varphi_\alpha$ and~$\varphi_\beta$ are close to~$e\in G$, hence we could glue them together to obtain a gauge~$\varphi_{\alpha\beta}$ on~$U_\alpha\cup U_\beta$ such that~$(\varphi_{\alpha\beta}^*\omega, \varphi_{\alpha\beta}^*\phi,\varphi^*_{\alpha\beta}\psi)$ is smooth throughout~$U_\alpha \cup U_\beta$.
	The detailed constructions can be found, for example, in~\cite{uhlenbeck1982connections} or~\cite{song2011critical}.
	Since there are only finitely many open sets in the cover, we obtain a global gauge~$\varphi\in\Gaug^{2,2}$ such that~$(\varphi^*\omega,\varphi^*(\phi),\varphi^*(\psi))$ is smooth.
\end{proof}

\section{Small energy regularity}%
\label{sec:SmallEnergyRegularity}
The small energy regularity result  for harmonic maps says that  the \(W^{2,2}\)-norm of the map is bounded by its Dirichlet energy on a slightly smaller domain.
This is a key estimate for establishing the energy identities, see e.g.~\cite{sacks1981existence}.
In this section we shall prove a small energy regularity result for the Yang--Mills--Higgs--Dirac model in preparation for the blow-up analysis in the following Section~\ref{sec:BlowUpAnalysis}

Since the Dirac action may be negative, which makes the action functional non-coercive, we have to use here another \emph{energy} of the spinorial fields.
More precisely, we introduce the following energies for the three fields in our model: for an open subset~$U\subset M$
\begin{align}
	\YM (\omega;U) = \int_U |F(\omega)|^2 \dv_g,& &
	E(\phi; U) = \int_U |\dd^\VB\phi|^2\dv_g, & &
	E(\psi; U) = \int_U |\psi|^4\dv_g,
\end{align}
and furthermore, by the notation~$E(\phi,\psi;U)$ we mean the sum of the energies of the section~$\phi$ and the spinor~$\psi$.
When~$U=M$, we will omit the domain if there is no confusion.
Due to the conformal invariance/covariance in dimension two, it is reasonable to have the smallness assumptions on small domains.
Thus we can restrict the model to a small disk where the bundles are trivialized.
For simplicity of notation we may assume that the local metric is Euclidean.

Let~$B$ be a Euclidean disk and consider the bundle~$P=B\times G$ with connection~$\omega$.
The associated bundle is~$\FB=B\times \fiber$, and the section is locally given by a map~$u\colon B\to N$.
The induced covariant derivative is as before given by
\begin{equation}%
\label{eq:local A-derivative of u}
	\dd_A u= \dd u+ \dd\mu_u(A).
\end{equation}
As a local map, the Dirichlet energy of~$u$ is
\begin{equation}
	E(u;B)=\int_B |\dd u|^2\dx.
\end{equation}
By~\eqref{eq:local A-derivative of u}, and up to a gauge if necessary, we have
\begin{equation}
	\left|\|\dd u\|_{L^2(B)}-\|\dd_A u\|_{L^2(B)}\right|
	\le C \|A\|_{L^2(B)}\le C \YM(A;B) .
\end{equation}
Thus, locally, we may not distinguish the classical Dirichlet energy of~$u$ with its vertical energy as a local section.

For later convenience, let us consider the approximating (local) system:
\begin{equation}%
\begin{split}%
\label{eq:EL-local-with errors}
	\hspace{.7cm}
	\dd^*\dd A
	={}&-\frac{1}{2}\dd^*[A,A]+A\llcorner\dd A
			+\frac{1}{2}A\llcorner [A,A]
			-{(\dd\mu_u)}^t\left(\dd u+ \dd\mu_u(A)\right)\\
		& -\left<\psi^j,\gamma(e_\alpha)\psi^i\right>
			\left<\partial_1\partial_2\bar{\mu}(\epsilon_a,\p_{y^i}),\p_{y^j}\right>
			\epsilon^a\otimes e_\alpha + \chi_1, \\
	\Delta u
	={}& \tr\II(u)(\dd u,\dd u)
			-2\tr\partial_1\partial_2\mu(A,\dd u)
			-\dd\mu_u(\diverg A)
			-\tr\partial_1\partial_2\mu_u \left(A,\dd\mu(A)\right) \\
		& +\frac{1}{2}\left<\psi^i,\gamma(e_\alpha)\psi^j\right>
			\left<\p_{y^i}, R\left(\p_{y^k}, \dd u(e_\alpha)
			+\dd\mu_u(A(e_\alpha))\right)\p_{y^j}\right> h^{kl}(u) u^*(\p_{y^{l}}) + \chi_2,\\
	\pd\psi^i
	={}&
		-\left\{- \Gamma^\eta_{\alpha k}(u){\left(\dd\mu_u A(\p_{x^\eta})\right)}^i+ \Gamma^i_{\alpha k}(u)+\Gamma^i_{jk}(u) u^j_\alpha\right\}\gamma(e_\alpha)\psi^k
		+\chi_3^i,
\end{split}
\end{equation}
with~$\chi_1,\chi_2,\chi_3$ being vector valued error terms such that
\begin{equation}
	\|\chi_1\|_{L^2(B)}^2
	+ \|\chi_2\|_{L^2(B)}^2+ \|\chi_3\|_{L^4(B)}^4 \le C<+\infty.
\end{equation}

\begin{prop}%
\label{prop:small energy regularity for A}
	Let~$(A,u,\psi)$ be a~$C^2$ solution of~\eqref{eq:EL-local-with errors} on~$B$.
	There exists an~$\varepsilon_0>0$ s.t.\ if
	\begin{align}
		\YM(A;B)\le \varepsilon_0,
	\end{align}
	then for any open disk~$B'\Subset B$, there exists~$C=C(B,B')>0$ such that
	\begin{align}
		\|A \|_{W^{2,2}}(B')
		\le& C\left(\YM(A;B)+ E(u;B)+E(\psi;B) \right)
			+C\|\chi_1\|_{L^2(B)}^2.
	\end{align}
\end{prop}

\begin{proof}
	By shifting the origin of the ambient space~$\R^K$, into which~$N$ is isometrically embedded, we may assume that the mean value~\(\bar{u}\) vanishes.
	Let~$B=U_1\Supset U_2 \Supset B' $.
	Let~$\epsilon_0<\kappa(2)$ so that we can apply a Coulomb gauge and assume that the statement in Theorem~\ref{thm:Uhlenbeck Coulomb gauge} holds.
	In particular~$\dd^* A=0$ and~$\|A\|^2_{W^{1,2}(B)}\le C\cdot \YM(A;B)$.

	Let~$\eta\in C^\infty_0(B) $ be a local cutoff function with~$\eta\equiv 1$ on~$U_2$.
	The local equation for~$\eta A$ is
	\begin{align}
	\Delta(\eta A)
	={}& (\Delta\eta)A+2\nabla\eta\cdot\nabla A
			+\frac{1}{2}\dd^*[\eta A,A]-\frac{1}{2}\dd\eta\lrcorner[A,A]
			-\eta A\lrcorner [A,A] \\
		& + {(\dd\mu_u)}^t\dd(\eta u) - {(\dd\mu_u)}^t(u\dd\eta)
			+ {(\dd\mu_u)}^t(\dd\mu_u)(\eta A) \\
		&+\left<\eta\psi^j,\gamma(e_\alpha)\psi^i\right>\left<\partial_1\partial_2\bar{\mu}(\epsilon_a,\p_{y^i}),\p_{y^j}\right>\epsilon^a\otimes e_\alpha
		-\eta\chi_1.
	\end{align}
	It follows that
	\begin{align}
	\|\Delta(\eta A)\|_{L^2(B)}
	\le& C(\eta)\|A\|_{L^2(B)} +C(\eta)\|\dd A\|_{L^2(B)}\\
		&+C\|\nabla(\eta A)\|_{L^4}\|A\|_{L^{4}(\supp \eta)}
		+C\|\eta A\|_{L^{\infty}(B)} \|\nabla A\|_{L^2(\supp\eta)}\\
		&+C(\eta)\|A\|_{L^{4}(B)}^2
			+ \|\eta A\|_{L^\infty(\supp\eta)}\|A\|_{L^{4}(\supp\eta)}^2 \\
		&+C(\mu,\fiber)\left( \|\dd(\eta u)\|_{L^2(B)}
		+C(\eta)\|u\|_{L^2(B)}+ \| A\|_{L^2(B)}\right) \\
		&+ C(\mu)\|\psi\|_{L^4(\supp\eta)}^2
			+ \|\chi_1\|_{L^2(B)}.
	\end{align}
	By the smallness assumptions on the energies, we have
	\begin{equation}\label{eq:small energy regularity for A}
	\|A\|_{W^{2,2}(U_2)}\le C\|\eta A\|_{W^{2,2}(B)}
	\le C\left(\| A\|_{W^{1,2}(B)}+ \|\nabla u\|_{L^2(B)}
	+ \|\psi\|_{L^4(B)}^2 + \|\chi_1\|_{L^2(B)}\right) ,
	\end{equation}
	where~$C=C(\mu,\fiber, \eta)>0$.
	Since~$\fiber$ and~$\mu$ are fixed, we actually have~$C=C(B,B')$.
\end{proof}

The small energy regularity for the other two fields can be obtained by employing the same strategy.
We omit the details.

\begin{prop}\label{prop:small energy regularity}
	Let~$(A,u,\psi)$ be a~$C^2$ solution of~\eqref{eq:EL-local-with errors} on~$B$.
	There exists an~$\varepsilon_0>0$ s.t.\ if
	\begin{align}
		\max\{\YM(A;B), E(u;B), E(\psi; B)\}\le \varepsilon_0,
	\end{align}
	then for any open disk~$B'\Subset B$, there exists~$C=C(B,B')>0$ such that
	\begin{align}
		\|u-\bar{u}\|^2_{W^{2,2}(B')} +\|\psi\|^2_{W^{1,4}(B')}
		+\|A \|_{W^{2,2}}(B')
		\le& C\left(\YM(A;B)+ E(u;B)+E(\psi;B) \right) \\
		& +C\left(\|\chi_1\|_{L^2(B)}^2+\|\chi_2\|_{L^2(B)}^2
			+\|\chi_3\|_{L^4(B)}^4\right).
	\end{align}
	Here~$\bar{u}$ is the mean value of~$u$ over~$B$.
\end{prop}

By the Sobolev embedding~$W^{2,2}(\R^2)\subset C^\beta_{loc}(\R^2)$, we get the following control on the oscillation of the section~$u$.
\begin{cor}
	Under the assumption of Proposition~\ref{prop:small energy regularity}, the oscillation of \(u\) is bounded by
	\begin{equation}
		\Osc_{B'} u \le
		C\left(\YM(A;B)+ E(u;B)+E(\psi;B) \right)
			+C\left(\|\chi_1\|_{L^2(B)}^2+\|\chi_2\|_{L^2(B)}^2
			+\|\chi_3\|_{L^4(B)}^4\right).
	\end{equation}
\end{cor}

If we can control the higher order derivatives of the error terms, then we can also control the higher order derivatives of the three fields under consideration.
In that case in the interior of the disk~$B$ the solutions with small energies are smoothly bounded.

\section{Blow-up analysis}%
\label{sec:BlowUpAnalysis}
In this section we investigate the compactness of the space of critical points of the Yang--Mills--Higgs--Dirac action functional.
As for Dirac-harmonic maps the coupled action does not satisfy the Palais--Smale condition for at least two reasons:
the maps part can form bubbles and the Dirac operator is non-definite.
In order to circumvent the second issue, we bound the \(L^4\)-norm or energy of the spinors together with the Dirichlet and Yang--Mills action.
Then we can establish that any sequence of approximating solutions contains a subsequence converging to a solution with possibly some bubbles.
The bubbles are particularly simple because the principal bundle and connection on them are trivial.

Let~$(\omega_k,\phi_k,\psi_k)$ be a sequence in the space~$\Conn^{1,2}\times W^{1,2}(\Gamma(\FB))\times W^{1,\frac{4}{3}}(\Gamma(S\otimes\phi^*\VB))$.
We say that \((\omega_k, \phi_k, \psi_k)\) is a sequence of approximating solutions to the Euler--Lagrange system~\eqref{eq:EL-global} if there are~$a_k \in L^2(\Gamma(\Ad(P)))$,~$b_k\in L^2(\Gamma(\phi_n^*\VB))$, and~$c_k\in L^4(\Gamma(S\otimes \phi_n^*\VB))$ such that
\begin{equation}%
\label{eq:small errors}
	\max (\|a_k\|_{L^2}, \|b_k\|_{L^2}, \|c_k\|_{L^4} ) \to 0 \quad
	\text{ as }\quad k\to \infty
\end{equation}
and
\begin{align}%
\label{eq:approximating system-global}
	&D_{\omega_k}^* F(\omega_k)+ \dd\bar{\mu}^*_{\phi_k}(\dd^\VB\phi_k)+ \Qpsi(\phi_k,\psi_k)= a_k ,\\
	&\tau^\VB(\phi_k)-\frac{1}{2}\VC(\phi_k,\psi_k)=b_k, \\
	&\D\psi_k= c_k.
\end{align}

\EnergyIdentitiesNoNeck

The proof of Theorem~\ref{thm:energy identities and no-neck} proceeds in several steps.
First we show that the Yang--Mills action does not concentrate in any point; hence up to subsequences the connections converge strongly.
Consequently no connection term appears in the bubbles (if a bubble exists); the bubbles are conformally invariant and the limits are Dirac-harmonic spheres.
Indeed, slightly stronger than Theorem~\ref{thm:energy identities and no-neck}, the points of energy concentration for the spinors are a subset of the concentration points for the maps.
Convergence away from the concentration points follows by the small energy regularity~\ref{prop:small energy regularity}.
In the last step it is necessary to consider the bubble formation by rescaling into an individual concentration point.

The remaining part is devoted to the proof of Theorem~\ref{thm:energy identities and no-neck}.
Most of the blowup analysis is well-developed and standard, see e.g.~\cite{jost2019energy} and the references therein; thus we only outline the argument, and emphasize the new technical issues.

\subsection{Proof of Theorem~\ref{thm:energy identities and no-neck}}
For the sequence \((\omega_k, \phi_k, \psi_k)\) of approximating solutions we define the following energy concentration sets
\begin{align}
	\mathcal{S}_1
	&\coloneqq \bigcap_{r>0} \left\{x\in M\mid \liminf_{k\to \infty} \int_{B_r(x)}|\dd^\VB\phi_k|^2 \dv_g \ge \varepsilon_0\right\}, \\
	\mathcal{S}_2
	&\coloneqq \bigcap_{r>0} \left\{x\in M\mid \liminf_{k\to \infty} \int_{B_r(x)}|\psi_k|^4 \dv_g \ge \varepsilon_0\right\}, \\
	\mathcal{S}_3
	&\coloneqq \bigcap_{r>0} \left\{x\in M\mid \liminf_{k\to \infty} \int_{B_r(x)}|F(\omega_k)|^2 \dv_g \ge \varepsilon_0\right\}.
\end{align}
As the energies are assumed to be uniformly bounded, each of the concentration sets consists of at most finitely many points or is possibly empty.

\begin{lemma}\label{lemma:no concentration for connection}
	$\mathcal{S}_3=\emptyset$.
\end{lemma}
\begin{proof}
	Suppose that there exists an~$x \in \mathcal{S}_3$.
	By passing to a subsequence we may assume that
	\begin{equation}\label{eq:concentration for A-by contradiction}
		\lim_{r\searrow 0}\lim_{k\to\infty}\int_{B_r(x)} |F(\omega_k)|^2\dv_g =\alpha(x)\ge \varepsilon_0.
	\end{equation}
	Choose~$0<r\ll 1$ so small that~$2r^2 \alpha(x)< \varepsilon_0$ and
	\begin{equation}
		\int_{B_r(x)} |F(\omega_k)|^2 \dv_g \le 2\alpha(x).
	\end{equation}
	Then, by rescaling via the map~$\theta_r\colon B_1(0)\to B_r(x)$ as in Lemma~\ref{lemma:scaling behaviors} we see that on~$B_1(0)$ the rescaled connections~${(\omega_k)}_r$ satisfy
	\begin{equation}
		\int_{B_1(0)} |F({(\omega_k)}_r)|^2\dx <\varepsilon_0.
	\end{equation}
	Then the estimate in Proposition~\ref{prop:small energy regularity for A} implies that, up to subsequences,~${(\omega_k)}_r$ converges strongly on~$B_1(0)$ in~$W^{1,2}$, say to~$\omega_\infty\in W^{2,2}(B_1(0))$.
	Scaling it back, we see that~$\omega_k$ converges strongly in~$W^{1,2}$ to~${(\omega_\infty)}_{1/r}$ on~$B_r(x)$, then again using Lemma~\ref{lemma:scaling behaviors}
	\begin{align}
		\lim_{r\searrow 0}\lim_{k\to\infty}\int_{B_r(x)} |F(\omega_k)|^2\dv_g
		&=\lim_{r\searrow 0}\int_{B_r(x)} |F({(\omega_\infty)}_{1/r})|^2\dv_g \\
		&=\lim_{r\searrow 0}r^{2}\int_{B_1(0)}|F(\omega_\infty)|^2\dv=0,
	\end{align}
	which contradicts the concentration inequality~\eqref{eq:concentration for A-by contradiction}.
\end{proof}

From Lemma~\ref{lemma:no concentration for connection} we see that the concentration set~$\mathcal{S}_1$ for the sections can be equivalently characterized by
\begin{equation}
	\mathcal{S}_1 =
	\left\{x\in M\mid \liminf_{k\to\infty}\int_{B_r(x)}|\dd u_k|^2 \dv \ge \varepsilon_0 \right\},
\end{equation}
since~$u$ has bounded values and the~$A_k$ part does not concentrate.

\begin{lemma}%
\label{lemma:sing-psi in sing-phi}
	$\mathcal{S}_2\subset \mathcal{S}_1$.
\end{lemma}
\begin{proof}
	Consider the local Dirac equation
	\begin{equation}
		\pd\psi^i_k
		=\left\{ \Gamma^\eta_{\alpha l}(u_k){\left(\dd\mu_{u_k} A_k(\p_{x^\eta})\right)}^i- \Gamma^i_{\alpha l}(u_k)-\Gamma^i_{jl}(u_k) {(u_k)}^j_\alpha\right\}\gamma(e_\alpha)\psi^l_k
			+\chi_{3k}^i,
	\end{equation}
	where~$c_k$ is locally represented by~$\chi_{3k}$.
	Taking a cutoff function~$\eta$ as before,
	we can localize the above equation as
	\begin{align}
		\pd(\eta\psi^i_k)
		= &\left\{\Gamma^\eta_{\alpha l}(u_k){\left(\dd\mu_{u_k} (\eta A_k)(\p_{x^\eta})\right)}^i
			- \eta\Gamma^i_{\alpha l}(u_k)
			-\Gamma^i_{jl}(u_k) {(\eta u_k)}^j_\alpha
			+\Gamma^i_{jl}(u_k) u_k^j\nabla_\alpha\eta\right\}\gamma(e_\alpha)\psi^l_k\\
			&+\gamma(\nabla\eta)\psi_k^i
			+\eta\chi_{3k}^i.
	\end{align}
	Then for any~$\frac{4}{3}<q<2$,
	\begin{align}
		\|\pd(\eta\psi_k)\|_{L^q(B_r(x))}
		\le& C\left(\|A_k\|_{L^2(B_r(x))} + \|\dd u_k\|_{L^2(B_r(x))} \right) \|\eta\psi_k\|_{L^{\frac{2q}{2-q}}(B_r(x))} \\
		& + \|\eta\psi_k\|_{L^q(B_r(x))}+ \|\chi_{3k}\|_{L^q(B_r(x))}.
	\end{align}

	If there was a point~$x\in \mathcal{S}_2\setminus \mathcal{S}_1$, then by taking~$r$ small, we may assume that
	\begin{equation}
		2C\left(\|A_k\|_{L^2(B_r(x))} + \|\dd u_k\|_{L^2(B_r(x))} \right) < C_q^{-1}
	\end{equation}
	with~$C_q$ being the Sobolev constant such that
	\begin{equation}
		\|\eta\psi_k\|_{W^{1,q}(B_r(x))} \le C_q\|\pd(\eta\psi_k)\|_{L^q(B_r(x))}.
	\end{equation}
	Then, shrinking~$r$ a little, we could control the~$W^{1,q}$ norm of~$\psi_k$ uniformly
	\begin{equation}
		\|\psi_k\|_{W^{1,q}(B_r(x))} \le
		C\left(\|\psi_k\|_{L^4(B_r(x))}+ \|\chi_{3k}\|_{L^4(B_r(x))}\right).
	\end{equation}
	Since the Sobolev embedding~$W^{1,q}\hookrightarrow L^4$ is compact in dimension two, it follows that, up to subsequences,~$(\psi_k)$ converges strongly in~$L^4(B_r(x))$ for~$r$ small.
	This contradicts the concentration assumption.
\end{proof}

	The small energy regularity, Proposition~\ref{prop:small energy regularity}, directly implies:
\begin{cor}\label{cor:convergence outside sing-phi}
	On~$M\setminus \mathcal{S}_1$, up to subsequences, the sequence~$(\omega_k,\phi_k,\psi_k)$ converges strongly.
\end{cor}

It remains to analyze the convergence near the finite set~$\mathcal{S}_1\equiv\{x_1, x_2, \dotsc, x_I\}$.
Note that, the weak limit~$(\omega_\infty, \phi_\infty, \psi_\infty)$, being itself a weak solution, is smooth by Theorem~\ref{thm:regularity of weak solution}.

As the blow-up procedure is purely local, we can restrict to a sufficiently small disk~$B_{\delta_i}(x_i)$ with fixed trivializations of the bundles.
Choose~$\delta_i>0$ small,~$1\le i\le I$, such that the balls~$B_{\delta_i}(x_i)$ are disjoint.
By passing to a subsequence we may assume that
\begin{equation}
	\lim_{\delta_i\searrow 0} \lim_{k\to \infty} \int_{B_{\delta_i}(x_i)} |\dd^\VB\phi_k|^2 \dv_g = \alpha(x_i)\ge \varepsilon_0.
\end{equation}

For simplicity of notation, we will assume that the Riemannian metric~$g$ on such a disk is Euclidean, while in the general case the metric may differ from the Euclidean metric by a small term if we employ geodesic normal coordinates.
The following proposition then implies Theorem~\ref{thm:energy identities and no-neck}

\begin{prop}%
\label{prop:blowup details}
	Let~$A_k\in \Conn^{1,2}$,~$u_k\in W^{1,2}(B_\delta(0),\fiber\subset\R^K)$, and~$\psi_k\in W^{1,\frac{4}{3}}(B_\delta(0),S\otimes u_n^*\R^K)$ be a sequence of solutions on the disk~$B_\delta(0)$ of the system~\eqref{eq:EL-local-with errors}, with uniformly bounded energies
	\begin{equation}
		E(A_k,u_k,\psi_k; B_\delta(0))
		=\int_{B_\delta(0)} |F_{A_k}|^2 +|\dd_{A_k} u_k|^2 +|\psi_k|^4\dx \le \Lambda <\infty,
	\end{equation}
	and the error term going to zero in norms
	\begin{equation}
		\|\chi_{1k}\|_{L^2}^2 +\|\chi_{2k}\|_{L^2}^2 +\|\chi_{3k}\|_{L^4}^4 \equiv \rho_k\to 0.
	\end{equation}
	Assume that they converge to~$(A_\infty, u_\infty, \psi_\infty)$ in~$W^{1,2}_{loc} \times W^{1,2}_{loc}\times W^{1,\frac{4}{3}}_{loc}(B_\delta(0)\setminus \{ 0\})$.
	Moreover assume~$0\in \mathcal{S}_1$, i.e.\ for any~$r>0$,
	\begin{equation}
		\liminf_{k\to\infty}\int_{B_r(0)} |\dd_{A_k} u_k|^2\dx \ge \varepsilon_0.
	\end{equation}
	Then there exists a positive integer~$I\in\mathbb{N}$ such that for each~$1\le i\le I$, there exist a sequence of points~$(x_k^i)\to 0$ and a sequence of small numbers~$\lambda_k^i \searrow 0$ such that
	\begin{enumerate}
		\item\label{item:separation} for any~$i \neq j$,
			\begin{equation}
				\frac{\lambda^i_k}{\lambda^j_k}
				+ \frac{\lambda^j_k}{\lambda^i_k}
				+ \frac{|x^i_n- x^j_n|}{\lambda^i_k + \lambda^j_k}
				=\infty;
			\end{equation}
		\item for each~$i$, the rescaled sequence
			\begin{align}
				\hat{A}_k^i(x)\coloneqq\lambda_k^i A_k(x^i_k+\lambda_k^i x) ,
				& &
				\hat{u}_k^i(x)\coloneqq u(x^i_k+\lambda_k^i x) , & &
				\hat{\psi}_k^i(x)\coloneqq \sqrt{\lambda_k^i}\psi_k(x^i_k+\lambda_k^i x),
			\end{align}
			converges to~$(0, \sigma^i, \xi^i)$in~$W^{1,2}_{loc}\times W^{1,2}_{loc}\times W^{1,\frac{4}{3}}_{loc}(\R^2)$, where~$(\sigma^i,\xi^i)$ extends to a Dirac-harmonic sphere.
		\item the energy identities hold:
			\begin{align}
				\lim_{k\to\infty}\YM(A_k;B_\delta(0)) &=\YM(A_\infty,B_\delta(0)), \\
				\lim_{k\to\infty} E(u_k;B_\delta(0))
				&=E(u_\infty;B_\delta(0))
					+\sum_{i=1}^I E(\sigma^i; \mathbb{S}^2), \\
				\lim_{k\to\infty} E(\psi_k; B_\delta(0))
				&=E(\psi_\infty;B_\delta(0))
					+\sum_{i=1}^I E(\xi^i;\mathbb{S}^2);
			\end{align}
		\item there is no neck between bubbles, i.e.\ the set~$u_\infty(B_\delta(0))\cup \left(\cup_{1\le i\le I} \sigma^i(\mathbb{S}^2)\right)$ is connected.
	\end{enumerate}
\end{prop}
\begin{proof}
	By passing to a subsequence if necessary, we assume that~$A_k$ converges to~$A_\infty$ in~$W^{1,2}$ strongly, and~$(u_k,\psi_k)$ converge to~$(u_\infty,\psi_\infty)$ weakly in~$W^{1,2}\times W^{1,\frac{4}{3}}(B_\delta(0))$ and strongly in~$W^{1,2}_{loc}\times W^{1,\frac{4}{3}}_{loc}(B_\delta(0)\setminus \{0\})$, with
	\begin{equation}
		\lim_{k\to\infty} \int_{B_r(0)} |\dd u_k|^2\dx \ge \varepsilon_0.
	\end{equation}
	The energy identity for the connections now follows.

	Let us construct and analyze the rescaling.
	Without loss of generality we consider the case that~$I=1$, i.e., there is only one bubble after rescaling, since there is a standard procedure to reduce the general situation to this case, see e.g.~\cite{ding1995energy}.
	Then we can drop the shoulder indices.

	For each~$k$, we choose~$\lambda_k>0$ such that
	\begin{equation}
		\sup_{x\in B_\delta(0)} E(u_k,\psi_k; B_{\lambda_k}(x))
		=\frac{\varepsilon_0}{4}
	\end{equation}
	and then choose~$x_k\in B_\delta(0)$ such that
	\begin{equation}
		E(u_k,\psi_k; B_{\lambda_k}(x_k))
		=\sup_{x\in B_\delta(0)} E(u_k,\psi_k; B_{\lambda_k}(x))
		=\frac{\varepsilon_0}{4}.
	\end{equation}
	By our assumption that the sequence converges strongly away from the origin, we conclude that~$|x_k|\to 0$ and~$\lambda_k\searrow 0$.
	The rescaled sequences are
		\begin{align}
			\hat{A}_k(x)\coloneqq\lambda_k A_k(x_k+\lambda_k x),
			& &
			\hat{u}_k(x)\coloneqq u(x_k+\lambda_k x), & &
			\hat{\psi}_k(x)\coloneqq \sqrt{\lambda_k}\psi_k(x_k+\lambda_k x),
		\end{align}
		which are defined on the ball~$B_{\delta/2\lambda_k}(0)\nearrow\R^2$ as~$k \to \infty$.
		From Lemma~\ref{lemma:scaling behaviors}, for an arbitrary~$R>1$,
		\begin{align}
			\YM(\hat{A}_k, B_R(0))
			= {(\lambda_k)}^2\YM(A_k;B_{\lambda_k R}(x_k))
			\le {(\lambda_k)}^2 \Lambda \to 0.
		\end{align}
		Moreover, the rescaled gauge potentials~$\hat{A}_k$ satisfy the equations:
		\begin{align}
			\dd^*\dd \hat{A}_k
			={}&-\frac{1}{2}\dd^*[\hat{A}_k,\hat{A}_k]+\hat{A}_k\llcorner\dd \hat{A}_k
					+\frac{1}{2}\hat{A}_k\llcorner [\hat{A}_k,\hat{A}_k]
					-\lambda_k^2{(\dd\mu_u)}^t\left(\dd \hat{u}_k+ \dd\mu_{\hat{u}_k}(\hat{A}_k)\right)\\
				& -\lambda_k^2\left<\hat{\psi}_k^j,\gamma(e_\alpha)\hat{\psi}_k^i\right>
					\left<\partial_1\partial_2\bar{\mu}(\epsilon_a,\p_{y^i}),\p_{y^j}\right>
					\epsilon^a\otimes e_\alpha
					+\lambda_{k}^3\hat{\chi}_{1k},
		\end{align}
		where~$\hat{\chi}_{1k}(x)\coloneqq \lambda_k \chi_{1k}(x_k+\lambda_k x)$ and~$\lambda_{1k}$ are the local representatives of~$a_k$.
		It follows that, up to Coulomb gauges,~$\hat{A}_k\to 0$ in~$W^{1,p}(B_\rho(x))$ for any~$B_\rho(0)\subset\R^2$ and any~$1<p<\infty$.
		Meanwhile~$(\hat{u}_k,\hat{\psi}_k)$ satisfies the system
		\begin{align}\label{eq:approximate DH-local-with errors}
			\Delta \hat{u}_k
			={}&\tr\II(\hat{u}_k)(\dd \hat{u}_k,\dd \hat{u}_k)
			+\frac{1}{2}\left<\hat{\psi}_k^i,\gamma(e_\alpha)\hat{\psi}_k^j\right>
					\left<\p_{y^i}, R\left(\p_{y^q},\dd\hat{u}_k(e_\alpha)\right)\p_{y^j}\right>
					h^{ql}(\hat{u}_k) {\hat{u}_k}^*(\p_{y^{l}})\\
				&-2\tr\partial_1\partial_2\mu(\hat{A}_k,\dd \hat{u}_k)
					-\dd\mu_{\hat{u}_k}(\diverg \hat{A}_k)
					-\tr\partial_1\partial_2\mu_{\hat{u}_k} \left(\hat{A}_k,\dd\mu_{\hat{u}_k}(\hat{A}_k)\right) \\
				& +\frac{1}{2}\left<\hat{\psi}_k^i,\gamma(e_\alpha)\hat{\psi}_k^j\right>
				\left<\p_{y^i}, R\left(\p_{y^q},\dd\mu_{\hat{u}_k}(\hat{A}_k(e_\alpha))\right)\p_{y^j}\right>
					h^{ql}(\hat{u}_k) {\hat{u}_k}^*(\p_{y^{l}})
					+ \hat{\chi}_{2k},\\
			\pd\hat{\psi}_k^i
			={}& \Gamma^i_{jl}(\hat{u}_k) {(\hat{u}_k)}^j_\alpha \gamma(e_\alpha)\hat{\psi}_k^l \\
				&+\left\{- \Gamma^\eta_{\alpha l}(\hat{u}_k){\left(\dd\mu_{\hat{u}_k} \hat{A}_k(\p_{x^\eta})\right)}^i+ \Gamma^i_{\alpha l}(\hat{u}_k)\right\}\gamma(e_\alpha)\hat{\psi}_k^l
					+\hat{\chi}_{3k}^i,
		\end{align}
		where~$\hat{\chi}_{jk}$ are the rescaling of~$\chi_{jk}$ ($j=2,3$), and~$\chi_{2k},\chi_{3k}$ are the local representative for~$b_k, c_k$.
		Their energies have the following quantitative properties:
		\begin{align}
			E(\hat{u}_k,\hat{\psi}_k; B_1(0))
			={}& E(u_k,\psi_k; B_{\lambda_k}(x_k))
				=\frac{\varepsilon_0}{4}, \\
			E(\hat{u}_k,\hat{\psi}_k; B_R(0))
			={}& E(u_k,\psi_k; B_{\lambda_k R}(x_k))
				\le \Lambda<\infty.
		\end{align}
		The system~\eqref{eq:approximate DH-local-with errors} shows that \((u_k, \psi_k)\) can be seen as a sequence of approximating solutions to the equations of Dirac-harmonic maps, see~\cite{jost2017blowup}.
		The error terms for the sections \(u_k\) are in~$L^2$ and the error terms for the spinors \(\psi_k\) are in~$L^4$, going to zero uniformly and scaling in the right way.
		Consequently, we can use the conclusion from~\cite{jost2017blowup} giving the convergence, bubbles, energy identities and no-neck statement.

		The proof of item~(\ref{item:separation}) is hidden in the reduction process on the number of bubbles.
		When blowing up, the rescaling parameter separates the concentration points; details can be found in e.g.~\cite{mcduff2012jholomorphic};
		This finishes the proof.
\end{proof}
This also completes the proof of Theorem~\ref{thm:energy identities and no-neck}.

Note that,  if~$x_0 \in\mathcal{S}_1$ but~$x_0\notin\mathcal{S}_2$, then the corresponding bubbles at~$x_0$ have trivial spinorial parts~$\xi^i\equiv 0$ and the mapping part~$\sigma^i$ defines a harmonic sphere.
This follows from the energy identity for spinors: if the blowup has a nontrivial spinorial part, there would correspond a energy concentration for spinors, contradicting that~$x_0\notin \mathcal{S}_2$.

\subsection{Concluding Remarks}
\begin{rmk}
	As in~\cite{jost2017blowup}, the proof of the blow-up actually gives
	\begin{equation}
		\lim_{k\to\infty} \int_{B_\delta(0)} |\nabla\psi_k|^{\frac{4}{3}}\dx
		=\int_{B_\delta(0)} |\nabla\psi_\infty|^{\frac{4}{3}}\dx
			+\sum_{i=1}^{I} \int_{\mathbb{S}^2} |\nabla\xi^i|^{\frac{4}{3}} \dv .
	\end{equation}
	Therefore we can get the global convergence of the action: in the notation of Theorem~\ref{thm:energy identities and no-neck}, denoting~$\omega_i^l\equiv 0$, so for each~$i$ and~$l$, the bundles are all trivial, and
	\begin{equation}
		\lim_{k\to\infty}\Action(\omega_k,\phi_k,\psi_k)
		=\Action(\omega_\infty,\phi_\infty,\psi_\infty)
		+ \sum_{i=1}^I \sum_{l=1}^{L_i} \Action(\omega_i^l, \sigma_i^l,\xi_i^l).
	\end{equation}
	As a corollary, if the fiber manifold~$(\fiber,\fbm)$ does not admit Dirac-harmonic spheres, then an approximating sequence with uniformly bounded energies must sub-converge to a smooth solution.
\end{rmk}
\begin{rmk}
	Consider the functional with a potential
	\begin{equation}
		\Action_\potential(\omega,\phi,\psi)
		=\int_M |F(\omega)|^2 + |\dd^\VB\phi|^2
			+ \left<\psi,\D\psi\right>+ \potential(\omega, \phi,\psi) \dv_g,
	\end{equation}
	where~$\potential(\omega,\phi,\psi)$ is~$G$-equivariant in the second variable and its derivatives satisfy
	\begin{align}
		|\potential_\omega| &\le C(1+|\dd_\omega\phi||\psi|^{\frac{s}{2}}+|\psi|^{s}), \\
		|\potential_\phi| &\le C(1+|\dd_\omega\phi||\psi|^{\frac{s}{2}}+|\psi|^{s}), \\
		|\potential_\psi| &\le C(1+|\dd_\omega\phi||\psi|^{\frac{s}{2m}}+ |\psi|^{s-1}),
	\end{align}
	for some~$s<\frac{2m}{m-1}$.
	The conditions on the potential imply that the perturbations caused by the potential are subcritical.
	The Euler--Lagrange system for this functional~$\Action_\potential$ is
	\begin{align}%
	\label{eq:EL-global-with potential}
		&D_\omega^* F+ \dd\bar{\mu}^*_\phi(\dd^\VB\phi)+ \Qpsi(\phi,\psi)+ \potential_\omega=0,\\
		&\tau^\VB(\phi)-\frac{1}{2}\VC(\phi,\psi)+\potential_\phi=0, \\
		&\D\psi+ \potential_\psi=0.
	\end{align}
	Under the above conditions on the potential the given proof for regularity of weak solutions in dimension two and the proof for the energy identities for approximating solutions with uniformly bounded energies generalize.
	The difficulty is to choose a potential that satisfies the mathematical constraints and is interesting from the viewpoint of physics or geometry.
\end{rmk}

\end{document}